\documentclass[aps,prl,twocolumn,showpacs,preprintnumbers]{revtex4}
\usepackage{amsmath}
\usepackage{graphicx}
\usepackage{dcolumn}
\usepackage{bm}
\usepackage{subfigure}
\usepackage{amsfonts}
\usepackage{amssymb}

\newcommand{\bmulticol}{\begin{multicols}{2}\narrowtext}
\newcommand{\emulticol}{\end{multicols}\widetext}

\begin{document}

\title{\textbf{Inverse kinetic theory for quantum hydrodynamic equations}}
\author{Massimo Tessarotto\thanks{%
Electronic-mail: Massimo.Tessarotto@cmfd.univ.trieste.it}$^{1,2},$ Marco
Ellero\thanks{%
Electronic-mail: Marco.Ellero@cmfd.univ.trieste.it}$^{3}$ \ and Piero
Nicolini\thanks{%
Electronic-mail: Piero.Nicolini@cmfd.univ.trieste.it}$^{1,2}$}
\affiliation{$^{1}$Dipartimento di Matematica e Informatica, Universit\`{a} di Trieste,
Italy\\
$^{2}$Consorzio di Magnetofluidodinamica, Trieste,Italy\\
$^{3}$Lehstuhl f\"{u}r Aerodynamik, Technische Universit\"{a}t
M\"{u}nchen, M\"{u}nchen, Germany}
\date{\today }

\begin{abstract}
We propose a solution for the inverse kinetic theory for quantum
hydrodynamic equations associated to the non-relativistic Schr\"{o}dinger
equation. It is shown that an inverse kinetic equation of the form of the
Vlasov equation can be non-uniquely determined under suitable mathematical
prescriptions.
\end{abstract}

\pacs{47.10.Fg,47.15.Km,11.10.Cd,11.15.Kc}
\maketitle


\section{I. Introduction}

A striking feature of standard quantum mechanics (SQM) is its analogy with
classical fluid dynamics. In fact, it is well-known that the Schr\"{o}dinger
equation \cite{Schroedinger1926} is equivalent to a closed set of partial
differential equations for suitable real-valued functions of position and
time (denoted as quantum fluid fields) \cite{Madelung1928}. The quantum
hydrodynamic description obtained in this way, which has been studied by
several authors \cite{Madelung1928,Van
Vleck1928,Moyal1949,Bohm1952a,Bohm1952b,Bohm1952c,Harvey,Schiller1952,Bohm1993,Holland,Deotto}%
, affords a straightforward physical interpretation in terms of a classical
fluid. This is obtained by identifying the classical fluid fields (number
density and fluid velocity) respectively with the quantum probability
density in configuration space and the quantum probability current (or
quantum fluid velocity). In particular, the quantum hydrodynamic equations
can be viewed as the equations of a classical compressible and non-viscous
fluid, endowed with potential velocity and quantized velocity circulation.

The analogy (with classical fluid dynamics) has motivated in the past
efforts to formulate phase-space techniques based on various statistical
models of quantum hydrodynamic equations. These works, although based on
different approaches, share the common view that the quantum state
corresponds to an underlying ("subquantum") statistical description of some
sort.

Starting from the pioneering work of Wigner \cite{Wigner1932,Bracken1999},
phase-space techniques typically require that the quantum fluid fields or
the wave-function itself be represented in terms of, or associated with,
appropriate phase-space functions. These are usually identified with
quasi-probabilities \cite{Cahill}, although formulations based on pure
probability densities are also available (see Ref.\cite{von Neumann}, sec.
IV.3). The procedure of representing quantum states by quasi-probabilities
in phase space is closely related to the phase space formulation of SQM
mechanics based on the noncommutative product known as Moyal product \cite%
{Groenwold,Moyal1949}. However, quasi-probabilities often become singular or
negative in the so-called "full quantum regime" \cite{Risken1988}, i.e.,
when the potential becomes nonlinear. This fact has motivated in the past
the search of alternative phase-space representations of the quantum state.
Among such approaches, we recall the phase space representation of SQM due
to Torres-Vega and Frederick \cite{Torres-Vega-1,Torres-Vega-2}, in which
the wavefunction is extended to phase space, $\psi (\Gamma ),$ and is
assumed to obey an appropriate "Schr\"{o}dinger equation in phase space".

Another class of statistical approaches is represented by the attempt to
interpret SQM in the framework of an underlying subquantum statistical
process. An example is provided by stochastic models (see for example Refs.
\cite{Nelson1966,Nelson1985,Kershaw1964,Namsray 1986,Blanchard1986} and the
review paper \cite{Bassi2003}), in which the underlying particle dynamics is
governed by stochastic differential equations, such as those describing the
nondissipative quantum Brownian motion, which traditionally is described by
Fokker-Planck or diffusion equations. Such equations, generally, lead again
to quasi-probability distributions which permit to "reconstruct" the Schr%
\"{o}dinger equation only approximately (namely in an asymptotic sense) and
under "ad hoc" initial conditions, since quasi-probability functions may
become, in general, invalid for non-Markovian processes with arbitrary noise
correlation. Although extensions of the classical theory of Brownian to
quantum domain have been proposed \ which permit the formulation in terms of
a well-behaved true probabilistic description \cite{Banik2005}, the problem
of these approaches remains that of explaining the origin of such
stochasticity \cite{Olavo-1,Olavo-2},\ which is precisely their weakness.

A second type of statistical approaches is provided by classical subquantum
kinetic models in which the underlying particle dynamics is purely
classical, i.e., the Schr\"{o}dinger equation is assumed to be based on
classical kinetic theory. An example of this type is provided by the
approach due to Kaniadakis \cite{Kaniadakis-2000-1,Kaniadakis-2003}, in
which each quantum particle is assumed to be composed by $N\gg 1$ identical
subquantum interacting classical particles (monads), to be described by
means of a classical kinetic equation. Also in this case the Schr\"{o}dinger
equation is - at best - recovered only in an approximate sense. Several
aspects of this formulation, however, remain unclear, which include - among
others - the problem of the closure of the moment equations, the
specification of the initial and boundary conditions for the kinetic
distribution function and the conditions of convergence to the Schr\"{o}%
dinger equation.

In this work we intend to investigate the problem of searching an \emph{%
inverse kinetic theory} for the Schr\"{o}dinger equation, i.e., a kinetic
theory yielding exactly, by means of a closed set of moment equations, the
quantum hydrodynamic equations. \ By definition, such a theory should be
non-asymptotic and able to yield the correct equations independently of the
initial and boundary conditions for the wavefunction, to be assumed suitably
regular. Furthermore, it should hold also for arbitrary, but suitably
smooth, initial conditions for the kinetic probability density and the
moment equations should form a complete system of equations, namely satisfy
a closure condition.

The construction of an inverse kinetic theory for the Schr\"{o}dinger
equation involves possibly also the identification of an underlying
classical dynamical system, in terms of which all relevant observables and
corresponding expectation values can in principle be advanced in time.

From the mathematical viewpoint the problem can be set, in principle, for
arbitrary fluid equations, an example being provided by the inverse kinetic
theory recently developed for the incompressible Navier-Stokes equation by
Ellero and Tessarotto \cite{Tessarotto2004,Ellero2005,Ellero2006} (hereon
referred to as ET). A basic prerequisite for the formulation of an inverse
kinetic theory of this type is, however, the proper definition of the
relevant \emph{quantum fluid fields }and their identification with suitable
momenta (to be denoted as \emph{kinetic fluid fields}), which include the
\emph{kinetic temperature} as well as the related definition of \emph{%
directional temperatures} (see below).

In this sense, the case of Schr\"{o}dinger equation is peculiar because, as
is well-known, its related fluid equations apparently depend only on two
quantum fluid fields, respectively, to be identified with the observables
quantum probability density and the quantum fluid velocity, while the
notions of quantum temperature and directional temperatures\ (to be
identified with the corresponding kinetic moments) remain in principle
arbitrary. In this paper we intend to propose a possible definition\ of
these observables which is related to the kinetic interpretation of the
Heisenberg theorem.

The problem is not merely of interest for theoretical and mathematical
research, but has potential relevance also for the understanding of the
fluid description of quantum mechanics and of the underlying statistical
models. Our motivation, in particular, is to exploit the analogy between
classical and quantum statistical mechanics, in order to prove that the
quantum observables and the fluid fields can formally be represented by
means of a purely classical statistical model. Although the mathematical
equivalence should not too hastily be regarded as implying physical
equivalence of the two formulations, this suggests that some relevant
classical reasonings can be transferred to SQM. This involves, in
particular, the formal description of SQM by means of a classical dynamical
system (to be denoted as \emph{phase-space Schr\"{o}dinger dynamical system}%
) which describes the dynamics of a set of virtual subquantum particles
interacting with each other only by means of an appropriate mean-field
interaction. \

Such a classical description is realized by means of an appropriate form of
the correspondence principle (denoted \emph{\ kinetic}), whereby the
physical observables, quantum fluid fields and quantum hydrodynamic
equations are respectively identified with appropriate ordinary phase-space
functions, velocity moments of the (subquantum) kinetic distribution
function and appropriate moment equations obtained from the inverse kinetic
equation. \

Here we intend to show, however, that in principle infinite solutions to
this problem exist, namely the inverse kinetic theory is intrinsically
non-unique. Such a feature is not surprising since these kinetic models may
be viewed simply as examples of the infinite admissible, and physically
equivalent, mathematical descriptions of physical reality. However, despite
such a non-uniqueness feature, it turns out that the functional form of the
kinetic equation and the mean-field force that defines the streaming
operator, by suitable prescriptions, can both be uniquely determined.

Another implication of the theory is to show in a simple way that the
Bohmian program, of reproducing the predictions of SQM within a framework in
which particles have definite Lagrangian trajectories \cite%
{Bohm1952a,Bohm1952b,Bohm1952c}, has by no means a unique solution, since
there are infinite equivalent possible realizations of the Schr\"{o}dinger
dynamical system which satisfy the quantum hydrodynamic equations. As is
well-known \cite{Holland,Deotto}, this means that there are in principle
infinite alternative theories to Bohmian Mechanics, which attribute
deterministic trajectories to the particle wave-function and are equivalent
to SQM.

The paper is organized as follows.

The mathematical setting of the hydrodynamic description of SQM is recalled
in Secs. II-IV. In particular, the quantum hydrodynamic equations for the
quantum fluid fields $\left\{ f,\mathbf{V}\right\} $\ are posed in Sec. II
and III, while the definition of the remaining quantum fluid fields,
provided by the quantum directional temperatures ($T_{QM,i},$ for $i=1,2,3$%
), is given in Sec.IV, together with their relationship with the Heisenberg
inequalities. The construction of the inverse kinetic theory is treated in
Secs.V and VI. In particular, the general form of the inverse kinetic
equation is laid in Sec.V, while in Sec.VI the mean-field force is
determined explicitly. The main results of theory are summarized in three
theorems which determine the explicit form of the inverse kinetic equation
and its basic properties.

\section{II. Hydrodynamic description of NRQM}

In this section we intend to recall the well-known fluid description of
non-relativistic quantum mechanics (NRQM), based on the property of the Schr%
\"{o}dinger equation to be equivalent to a complete set of fluid equations.
For the sake of clarity let us introduce the basic definitions and the
mathematical formulation of the problem.

In this paper we shall consider, in particular, the case of a system of
spinless scalar particles (bosons) described by a single scalar wavefunction
$\psi (\mathbf{r},t),$with associated probability density%
\begin{equation}
f=\left\vert \psi (\mathbf{r},t)\right\vert ^{2},
\label{probability density}
\end{equation}%
requiring that both are defined and continuous in $\overline{\Omega }\times
I $ and in addition that $f$ is strictly positive in $\Omega ,$ $\overline{%
\Omega }$ denoting the closure of $\Omega .$ Hence, by assumption, $f$\ $\ $%
can only vanish on the boundary $\delta \Omega $ (i.e., in the nodes $%
\mathbf{r}_{n}\in \delta \Omega $ where $f(\mathbf{r}_{n},t)=0$) and must
satisfy the normalization
\begin{equation}
\int_{\Omega }d\mathbf{r}f(\mathbf{r},t)=1.  \label{normalization for f}
\end{equation}%
In addition, $f$\ and $\psi $\ are respectively assumed single-valued and
possibly multi-valued in $\Omega \times I,$\ with $\psi $\ at least of class
$C^{(3,2)}(\Omega \times I)$. For definiteness, we shall also assume,
without loss of generality, that $\Omega $\ is a connected subset of $%
\mathbb{R}
^{3N}$ and $\psi (\mathbf{r},t)$ belongs to the functional space $\left\{
\psi \right\} ,$ to be identified with the Hilbert space of complex-valued
functions which are square-integrable in $\Omega $. The $N-$body
wave-function $\psi (\mathbf{r},t)$ is required to satisfy in the open set $%
\Omega \times I$ the vector\emph{\ }Schr\"{o}dinger equation
\begin{equation}
i\hbar \frac{\partial }{\partial t}\psi =H\psi ,  \label{Eq.3b}
\end{equation}%
where $H=H_{o}+U$\ is the $N-$body Hamiltonian operator. Here, denoting $%
\nabla _{(j)}\equiv \frac{\partial }{\partial \mathbf{r}_{j}},$\ $%
H_{o}=-\sum\limits_{j=1,N}\frac{\hbar ^{2}}{2m}\nabla _{j}^{2}$ and $U$ are
respectively the free-particle Hamiltonian (kinetic energy) and the
interaction potential, to be identified with a real function defined and
suitably smooth in $\Omega \times I.$\ \ For well-posedness, appropriate
initial and boundary conditions must be imposed on $\psi (\mathbf{r},t)$.
The initial conditions are obtained by imposing for all $\mathbf{r}\in
\overline{\Omega }$\
\begin{equation}
\psi (\mathbf{r},t_{o})=\psi _{o}(\mathbf{r}),  \label{Eq.4a}
\end{equation}%
where $\psi _{o}$\ is a suitably smooth complex-valued function. To specify
the boundary conditions, we first notice the boundary set $\delta \Omega $
can always be considered prescribed. The boundary conditions can be
specified by imposing Dirichlet boundary conditions on $\delta \Omega $.
This requires $\forall \mathbf{r}_{\delta }\in \delta \Omega $

\begin{equation}
\psi (\mathbf{r}_{\delta },t)=\psi _{w}(\mathbf{r}_{\delta },t),
\label{b.c.for f - prescribed}
\end{equation}%
\begin{equation}
\lim_{\mathbf{r}\rightarrow \mathbf{r}_{\delta }}\mathbf{V}(\mathbf{r},t)=%
\mathbf{V}_{w}(\mathbf{r}_{_{\delta }},t),  \label{b.c. for V - prescribed}
\end{equation}%
where $V(\mathbf{r},t)$ is the quantum velocity field
\begin{equation}
\mathbf{V}(\mathbf{r},t)=\frac{\hbar }{2mi\left\vert \psi (\mathbf{r}%
,t)\right\vert ^{2}}\left\{ \psi ^{\ast }\nabla \psi -\psi \nabla
\psi^{\ast } \right\} .  \label{quantum velocity}
\end{equation}%
Here the complex function $\psi _{w}(\mathbf{r}_{_{\delta }},t)$ and the
real vector function $\mathbf{V}_{w}(\mathbf{r}_{_{\delta }},t)$ are
prescribed, suitably smooth functions. To specify the value of $f(\mathbf{r}%
,t)$ on $\delta \Omega $, let us require that there results additionally%
\begin{equation}
\int_{\Omega }d\mathbf{r}\nabla f(\mathbf{r},t)=0.
\label{boundness condition}
\end{equation}%
In all such cases Eq.(\ref{boundness condition}) implies that there must be $%
\forall \mathbf{r}_{_{\delta }}\in $ $\delta \Omega $
\begin{equation}
f(\mathbf{r}_{_{\delta }},t)=\left\vert \psi _{w}(\mathbf{r}_{_{\delta
}},t)\right\vert ^{2}\equiv f_{o}\geq 0,  \label{b.c. for f - 1st case}
\end{equation}%
where $f_{o}$ is either a constant, whose value may still depend on the
specific subset$,$ or at most is a function $f_{o}(t)$ to be assumed
suitably smooth $\forall t\in $ $I.$ Hence, the points of $\delta \Omega $
are not necessarily nodes. However, if $\mathbf{r}_{_{\delta }}$ is an
improper point of $%
\mathbb{R}
^{3N}$ (hence, $\Omega $ is assumed to be an unbounded subset of $%
\mathbb{R}
^{3N}$), since it must be $\lim_{\left\vert \mathbf{r}\right\vert
\rightarrow \infty }f(\mathbf{r},t)=0,$ $\mathbf{r}_{_{\delta }}$ is
necessarily a node, i.e.,
\begin{equation}
f_{o}=0.  \label{b.c. for f -2nd case}
\end{equation}%
This implies for consistency also
\begin{equation}
\lim_{\left\vert \mathbf{r}\right\vert \rightarrow \infty }\psi _{w}(\mathbf{%
r},t)=0.  \label{b.c. for psi-2nd case}
\end{equation}

The set of equations (\ref{Eq.3b}),(\ref{Eq.4a}),(\ref{b.c.for f -
prescribed}),(\ref{b.c. for V - prescribed}) together with (\ref{b.c. for f
- 1st case}) or (\ref{b.c. for f -2nd case}) and (\ref{b.c. for psi-2nd case}%
), defines the initial-boundary value problem for the Schr\"{o}dinger
equation (\emph{SE problem}). The solution of the SE problem, $\psi ,$ must
be determined in an appropriate functional space, to be suitably defined
(see for example Ref.\cite{Gustafson-Sigal2005}).

The set of hydrodynamic equations corresponding to the Schr\"{o}dinger
equation are well-known \cite{Madelung1928,Van
Vleck1928,Schiller1952,Bohm1952a} and follow immediately from the
exponential representation (known as \emph{Madelung transformation} \cite%
{Madelung1928})
\begin{equation}
\psi =\sqrt{f}e^{i\frac{S}{\hbar }},  \label{Eq.Madelung}
\end{equation}%
where $\left\{ f,S\right\} ,$ denoted as \emph{quantum fluid fields, }are
respectively the \emph{quantum probability density} and the \emph{quantum
phase-function} (also denoted as Hamilton-Madelung principal function). Eq. (%
\ref{Eq.Madelung}) is manifestly defined only in the set in which results $%
f>0$ (i.e., in the configuration space $\Omega $)$.$ We stress that in
principle $S(\mathbf{r},t)$\ remains "a priori" unspecified on the subset
the boundary $\delta \Omega $ where $f=0$ (subset of the nodes $\mathbf{r}%
_{n}$)$.$ This indeterminacy, however, is eliminated by requiring that
everywhere in $\delta \Omega ,$ $S(\mathbf{r},t)$ can be prolonged on the
same set by imposing $\forall \mathbf{r}_{n}\in \delta \Omega :$%
\begin{equation}
S(\mathbf{r}_{n},t)\equiv \lim_{\mathbf{r\rightarrow }\mathbf{r}_{n}}S(%
\mathbf{r},t).  \label{limit-nodes}
\end{equation}%
Hence, the real functions $\left\{ f,S\right\} $ can both be assumed
continuous in $\overline{\Omega }\times I$ and at least $C^{(3,1)}(\Omega
\times I).$ Obviously, $S(\mathbf{r},t)$ is defined up to an additive
constant $2\pi k,$ being $k\in
\mathbb{Z}
,$ while $S$ itself is generally not single-valued. In addition, if $\psi $
is single-valued, it is obvious that $S$ must satisfy a well-defined
condition of multi-valuedness. In fact, in this case on any regular closed
curve $C$ of $\Omega $, $S$ it must result
\begin{equation}
\int\limits_{C}d\mathbf{l\cdot \nabla }S(\mathbf{r},t)=2\pi n\hbar ,
\label{condition of multi-valuedness}
\end{equation}%
where $n$ is an appropriate relative number \cite{Walstrom}. Introducing the
single-valued potential velocity field, defined in $\Omega \times I,$%
\begin{equation}
\mathbf{V}(\mathbf{r},t)=\frac{1}{m}\nabla S(\mathbf{r},t),
\label{potntial velocity}
\end{equation}%
this yields the well-known condition of quantization of the velocity
circulation%
\begin{equation}
\kappa \equiv \int\limits_{C}d\mathbf{l\cdot V}(\mathbf{r},t)=\frac{2\pi
n\hbar }{m}.  \label{quantization}
\end{equation}%
Hence, by denoting $\frac{D}{Dt}=\frac{\partial }{\partial t}+\mathbf{V\cdot
\nabla }$ (convective derivative), it follows that in the open domain $%
\Omega \times I$ (where by definition $f>0$) the fluid fields $\left\{
f,S\right\} $ obey the complete set of hydrodynamic equations represented
respectively by
\begin{eqnarray}
&&\left. \frac{Df}{Dt}+f\nabla \cdot \mathbf{V}=0,\right.  \label{Eq.10} \\
&&\left. \frac{\partial S}{\partial t}+\frac{1}{2m}\left\vert \nabla
S\right\vert ^{2}=-U_{QM}\right. .  \label{Eq.11}
\end{eqnarray}%
These are denoted as \emph{quantum hydrodynamic equations}. The first one is
manifestly the continuity equation for the quantum probability density $f(%
\mathbf{r},t).$ Instead, the second one is the Hamilton-Jacobi equation for
the quantum phase-function $S.$ Moreover, $U_{QM}$ is the so-called quantum
potential \cite{Bohm1952a} related to the interaction potential $U$ by means
of the equation $U_{QM}=-\frac{\hbar ^{2}}{2}\left( \frac{1}{2}\nabla
^{2}\ln f+\frac{1}{4}\left\vert \nabla \ln f\right\vert ^{2}\right) +U.$
Since by assumption $U$ is single-valued in $\Omega $, it follows that
equations (\ref{Eq.10}) and (\ref{Eq.11}) must also be single-valued.
Nevertheless, both $\frac{\partial S}{\partial t}$ and $U_{QM}$ are not
unique since they are determined up to an arbitrary real smooth function $%
z(t)$ since they are invariant with respect to the gauge transformation
\begin{equation}
\left\{
\begin{array}{c}
S(\mathbf{r},t)\rightarrow S^{\prime }(\mathbf{r},t)=S(\mathbf{r},t)+\frac{1%
}{\hbar }\int dt^{\prime }z(t^{\prime }), \\
U(\mathbf{r},t)\rightarrow U^{\prime }(\mathbf{r},t)=U(\mathbf{r},t)+z(t),
\\
f(\mathbf{r},t)\rightarrow f^{\prime }(\mathbf{r},t)=f(\mathbf{r},t).%
\end{array}%
\right\}  \label{Eq.15}
\end{equation}

The initial conditions to be satisfied by the quantum fluid fields $\left\{
f,S\right\} ,$ stem from Eq. (\ref{Eq.4a}) and read:

\begin{equation}
f(\mathbf{r},t_{o})=f_{o}(\mathbf{r}),  \label{initial condition for f}
\end{equation}%
\begin{equation}
S(\mathbf{r},t_{o})=S_{o}(\mathbf{r})\text{\textbf{\ }}mod(\mathbf{2}\pi ).
\label{initial condition for V-1}
\end{equation}%
Instead, the boundary conditions implied by Eqs. (\ref{Eq.3b}),(\ref{Eq.4a}%
),(\ref{b.c.for f - prescribed}),(\ref{b.c. for V - prescribed}) together
with (\ref{b.c. for f - 1st case}) or (\ref{b.c. for f -2nd case}) and (\ref%
{b.c. for psi-2nd case}), read respectively $\forall \mathbf{r}_{\delta }\in
\delta \Omega $
\begin{equation}
f(\mathbf{r}_{\delta },t)=f_{w}(\mathbf{r}_{\delta },t)
\label{BS f -PRESCRIBED}
\end{equation}%
\begin{equation}
S(\mathbf{r}_{\delta },t)=S_{w}(\mathbf{r}_{\delta },t)\text{ }mod(\mathbf{2}%
\pi )  \label{BC S - PRESCRIBED}
\end{equation}%
\begin{equation}
\lim_{\mathbf{r}\rightarrow \mathbf{r}_{\delta }}\mathbf{V(r},t)=\mathbf{V}%
_{w}(\mathbf{r}_{\delta },t),  \label{BC V - PRESCRIBED -1}
\end{equation}%
where $S_{w}(\mathbf{r}_{\delta },t)$ and $\mathbf{V}_{w}(\mathbf{r}_{\delta
},t)$ are suitably smooth real functions and $f_{w}(\mathbf{r}_{\delta },t)$
is specified either by Eq.(\ref{b.c. for f - 1st case}) or (\ref{b.c. for f
-2nd case}), depending on the definition of the domain $\Omega $.

Equations (\ref{Eq.10}),(\ref{Eq.11}), together with the initial conditions (%
\ref{initial condition for f}),(\ref{initial condition for V-1}) and the
boundary conditions (\ref{BS f -PRESCRIBED})-(\ref{BC V - PRESCRIBED -1}),
define the quantum hydrodynamic initial-boundary problem (\emph{QHE problem}%
).

\section{III. Gauge-invariant form of the hydrodynamic equations}

The gauge function $z(t)$ can be eliminated by taking the gradient of Eq.(%
\ref{Eq.11}) term by term. The resulting gauge-independent equations for the
quantum fluid fields $\left\{ f,\mathbf{V}\right\} ,$ again valid in the
open domain $\Omega \times I,$ are provided by the \emph{gauge-invariant
quantum hydrodynamic equations, }which are defined by the continuity
equation (\ref{Eq.10}) and by \
\begin{equation}
\frac{D}{Dt}\mathbf{V}(\mathbf{r},t)=\frac{1}{m}\mathbf{F}\equiv -\frac{1}{m}%
\nabla U_{QM}.  \label{eq.11c}
\end{equation}%
As a consequence, by Eqs.(\ref{Eq.10}) and (\ref{eq.11c}) can be viewed as
the \emph{hydrodynamic equations of a compressible fluid}. \ On the other
hand, Eq.(\ref{eq.11c}) implies%
\begin{equation}
\frac{\partial }{\partial t}\nabla \times \mathbf{V}(\mathbf{r},t)+\nabla
\times \left( \mathbf{V\cdot \nabla V}(\mathbf{r},t)\right) =0,
\end{equation}%
where $\mathbf{V\cdot \nabla V}=-\mathbf{V\times }\left( \mathbf{\nabla
\times V}\right) -\nabla V^{2}$ and%
\begin{eqnarray}
&&\left. \nabla \times \left( \mathbf{V\cdot \nabla V}(\mathbf{r},t)\right)
=-\nabla \times \left[ \mathbf{V\times \xi }\right] =\right. \\
&&\left. =-\mathbf{\xi }\cdot \nabla \mathbf{V+V}\cdot \nabla \mathbf{\xi }%
\right. .  \notag
\end{eqnarray}%
where $\mathbf{\xi =\nabla \times V}$ is the vorticity vector. Therefore, if
we impose in the whole domain $\Omega $ the initial condition $\mathbf{\xi
(r,}t_{o})=\mathbf{0}$ it results
\begin{equation}
\mathbf{\xi (r,}t)=\mathbf{0}  \label{zero vorticity}
\end{equation}%
for all $t\in I.$\ Notice that Eq.(\ref{zero vorticity}) is not in
contradiction with the condition of quantization for the velocity
circulation $\kappa $\ \ [see Eq.(\ref{quantization})] since the phase
function $S(\mathbf{r},t)$ results generally non-single-valued. As a
consequence, the vector field $\mathbf{V}(\mathbf{r},t)$ is necessarily of
the form (\ref{potntial velocity}). Hence, the fluid described by the fluid
fields $\left\{ f,\mathbf{V=}\frac{1}{m}\nabla S(\mathbf{r},t)\right\} $ is
necessarily \emph{vorticity-free, }while at the same time\emph{\ }its
velocity circulation is non-vanishing [see Eq.(\ref{quantization})]. This
equation is know as the so-called \emph{quantum Newton equation\ }\cite%
{Bohm1952a}(or \emph{quantum Navier-Stokes equation} \cite{Harvey}). The
initial-boundary conditions for these equations are defined again by (\ref%
{initial condition for f}),(\ref{initial condition for V-1}) and (\ref{BS f
-PRESCRIBED})-(\ref{BC V - PRESCRIBED -1}), which imply in particular for $%
\mathbf{V}(\mathbf{r},t_{o})$ the initial condition:

\begin{equation}
\mathbf{V}(\mathbf{r},t_{o})=\frac{1}{m}\nabla S_{o}(\mathbf{r,}t_{o}).
\label{initia c-V}
\end{equation}

Equations (\ref{Eq.10}),(\ref{eq.11c})], together with the initial
conditions (\ref{initial condition for f}),(\ref{initial condition for V-1})
and the boundary conditions (\ref{BS f -PRESCRIBED})-(\ref{BC V - PRESCRIBED
-1}), define the gauge-invariant quantum hydrodynamic initial-boundary
problem (\emph{GI-QHE problem}).

In summary, by construction it follows that:

a) the QHE problem is equivalent to the SE problem, namely $\left\{
f,S\right\} $ is a solution of the first problem if and only if $\psi (%
\mathbf{r},t)$ is a solution of the second one; as a consequence the
solution $\left\{ f,S\text{ }mod(2\pi )\right\} $ of the QHE problem is
unique;

b) if $\left\{ f,S\text{ }mod(2\pi )\right\} $ is a solution of the QHE
problem then $\left\{ f,\mathbf{V=}\frac{1}{m}\nabla S(\mathbf{r},t)\right\}
$ is necessarily a solution of the GI-QHE problem;

c) vice versa, a solution $\left\{ f,\mathbf{V}\right\} $ of the GI-QHE
problem, determines uniquely $\left\{ f,S\right\} $ up to an arbitrary gauge
transformation of the form (\ref{Eq.15}).

Finally, it is worthwhile to mention that, in principle, it is also possible
to introduce sets of "reduced" hydrodynamic equations, defined in
appropriate subspaces of the $N-$body configuration space, in particular the
one-particle subspaces $\Omega _{i}$ (for $i=1,N$). The latter can be
obtained adopting for the $N-$body quantum system the one-particle reduced
representation described in the Appendix A [see Eq.(\ref{one-particle reduce
representation})]. Manifestly, these reduced descriptions are not equivalent
to the full $N-$body description. For example, the $N-$body system can be
considered as formed by $N$ $\ 1$-body subsystem, one for each particle ($%
j=1,N$). For each 1-body subsystem it is possible to introduce a set of
reduced quantum fluid fields $\left\{ f_{j},\mathbf{V}_{j}\mathbf{=}\frac{1}{%
m}\nabla _{j}S_{j}(\mathbf{r}_{j},t)\right\} ,$ both defined in the set $%
\Omega _{j}\times I$ and uniquely associated to the one-particle wave
function $\psi _{j}(\mathbf{r}_{j},t)$ by means of Eqs. (\ref{probability
density}) and (\ref{potntial velocity}). It is immediate to prove that the
fluid fields $\left\{ f_{j},\mathbf{V}_{j}\right\} $ for $j=1,N$ obey a set
of fluid equations formally analogous to Eqs. (\ref{Eq.10}),(\ref{eq.11c}),
to be denoted as \emph{reduced hydrodynamic equations}, which can be viewed
as describing the dynamics of an \emph{immiscible fluid mixture}.

\section{IV. Heisenberg theorem and the concept of quantum temperature}

The set fluid equations Eqs. (\ref{Eq.10}) and (\ref{eq.11c}) for the
quantum fluid fields $\left\{ f,\mathbf{V}\right\} $ provide a complete
description of quantum systems. \ This means, in particular, that \emph{no
other} independent observable or fluid field is required to describe the
quantum state. However, for the subsequent analysis it is useful to
introduce the concepts of quantum directional temperatures and quantum
temperature, which can be defined by analogy with classical statistical
mechanics and interpreted as additional quantum fluid fields. We intend to
show that the definitions of these observables emerge from Heisenberg
theorem. We recall that this is realized by means of the (Heisenberg)
inequalities (holding for $i=1,2,3$)
\begin{equation}
\left\langle \left( \Delta r_{i}\right) ^{2}\right\rangle \left\langle
\left( \Delta p_{i}\right) ^{2}\right\rangle \geq \frac{\hbar ^{2}}{4},
\label{Heisenberg-inequalities}
\end{equation}%
or%
\begin{equation}
\overline{\Delta }r_{i}\overline{\Delta }p_{i}\geq \frac{\hbar }{2}.
\label{Heisenber inequalities-2}
\end{equation}%
Here\emph{\ }the notation is standard. Thus, $\overline{\Delta }%
r_{i}=\left\langle \left( \Delta r_{i}\right) ^{2}\right\rangle ^{1/2},%
\overline{\Delta }p_{i}=\left\langle \left( \Delta p_{i}\right)
^{2}\right\rangle ^{1/2}$ \ (for $i=1,2,3$) are the \emph{quantum standard
deviations }for\emph{\ }position and linear momentum, \ $\left\langle \left(
\Delta r_{i}\right) ^{2}\right\rangle ,\left\langle \left( \Delta
p_{i}\right) ^{2}\right\rangle $ are the corresponding \emph{average
quadratic quantum fluctuations, }while $\Delta r_{i},\Delta p_{i}$ are
respectively the \emph{quantum position and momentum fluctuations} $\Delta
r_{i}=r_{i}\mathbf{-}\left\langle r_{i}\right\rangle ,$ $\Delta p_{i}=p_{i}%
\mathbf{-}\left\langle p_{i}\right\rangle .$ Finally $\left\langle
Q\right\rangle \equiv \left\langle \psi \mid Q\psi \right\rangle
=\int_{\Omega }d\mathbf{r}f(\mathbf{r},t)Q(\mathbf{r},t)$ denotes the
expectation value of a generic dynamical variable $Q$. As usual we identify
the quantum linear momentum $\mathbf{p}$ with the linear differential
operator
\begin{equation}
\mathbf{p}=-i\hbar \nabla  \label{quantum linear momentum}
\end{equation}%
which acts on the functional space $\left\{ \psi \right\} .$ It follows $\
\left\langle \mathbf{p}\right\rangle \equiv \left\langle \psi \mid \mathbf{p}%
\psi \right\rangle =m\int_{\Omega }d\mathbf{r}f\mathbf{V}\equiv \left\langle
\mathbf{P}\right\rangle ,$ where
\begin{equation}
\mathbf{P}=m\mathbf{V}  \label{fluid momentum}
\end{equation}%
is the \emph{fluid momentum}, while the expectation value of $\ \left(
\Delta p_{i}\right) ^{2}$, upon integration on the set $\Omega $, reads
\begin{equation}
\left\langle \left( \Delta p_{j}\right) ^{2}\right\rangle =\frac{\hbar ^{2}}{%
4}\int\limits_{\Omega }d\mathbf{r}f\left( \partial _{j}\ln f\right)
^{2}+\left\langle \left( \partial _{j}S\right) ^{2}\right\rangle
-\left\langle \partial _{j}S\right\rangle ^{2}.  \label{position delta_pi}
\end{equation}

As is well-known Heisenberg theorem follows by invoking the identity%
\begin{equation}
\int\limits_{\Omega }d\mathbf{r}f(\mathbf{r},t)=\int\limits_{\Omega }d%
\mathbf{r}\left[ r_{i}\mathbf{-}\left\langle r_{i}\right\rangle \right]
\cdot \frac{\partial }{\partial r_{i}}f(\mathbf{r},t)=1,
\end{equation}%
which implies
\begin{equation}
\int\limits_{\Omega }d\mathbf{r}\sqrt{f\left[ r_{i}\mathbf{-}\left\langle
r_{i}\right\rangle \right] ^{2}}\sqrt{f\left[ \frac{\partial }{\partial r_{i}%
}\ln f(\mathbf{r,v},t)\right] ^{2}}\geq 1.
\end{equation}%
Hence, Schwartz inequality delivers:%
\begin{equation}
\left\langle \left( \Delta r_{i}\right) ^{2}\right\rangle
\int\limits_{\Omega }d\mathbf{r}f\left[ \frac{\partial }{\partial r_{i}}\ln
f(\mathbf{r,v},t)\right] ^{2}\geq 1,  \label{Schwartz
inequality}
\end{equation}%
where by definition $\left\langle \left( \Delta r_{i}\right)
^{2}\right\rangle \equiv \int\limits_{\Omega }d\mathbf{r}f\left[ r_{i}%
\mathbf{-}\left\langle r_{i}\right\rangle \right] ^{2}$.

A peculiar aspect of the\ Heisenberg inequality (\ref%
{Heisenberg-inequalities}) is that it can also be written in terms of the
relative fluctuations $\Delta ^{(1)}p_{i}=p_{i}-P_{i},$ which are \emph{%
defined with respect to the components of fluid momentum} $P_{i}=mV_{i}$
(for $i=1,2,3$)$,$ \emph{instead of the corresponding expectation values} $%
\left\langle P_{i}\right\rangle $. The analogy is based on the fact that the
concept of kinetic temperature, and the related one of directional kinetic
temperatures [see Eq.(\ref{moment -3A}) in the next Section], is defined in
terms of fluctuations with respect to the local fluid velocity $\mathbf{V(r,}%
t)$, instead its expectation value $\left\langle \mathbf{V(r,}%
t)\right\rangle .$ In fact, it is immediate to prove that, by definition of
the quantum linear momentum (\ref{quantum linear momentum}), the following
identity holds (for $i=1,2,3$)%
\begin{equation}
\left\langle \left( \Delta p_{i}\right) ^{2}\right\rangle =\left\langle
\left( \Delta ^{(1)}p_{i}\right) ^{2}\right\rangle +\left\langle \left(
\Delta ^{(2)}p_{i}\right) ^{2}\right\rangle ,
\label{momentum fluctuations-2}
\end{equation}%
where $\left\langle \left( \Delta ^{(1)}p_{i}\right) ^{2}\right\rangle $ and
$\left\langle \left( \Delta ^{(2)}p_{i}\right) ^{2}\right\rangle $ read
respectively
\begin{eqnarray}
\left\langle \left( \Delta ^{(1)}p_{i}\right) ^{2}\right\rangle &=&\frac{%
\hbar ^{2}}{4}\int\limits_{\Omega }d\mathbf{r}f\left( \partial _{j}\ln
f\right) ^{2},  \label{part A} \\
\left\langle \left( \Delta ^{(2)}p_{i}\right) ^{2}\right\rangle &\equiv
&\left\langle \left( \partial _{j}S\right) ^{2}\right\rangle -\left\langle
\partial _{j}S\right\rangle ^{2},  \label{part B}
\end{eqnarray}%
and hence can be interpreted as the average quadratic momentum fluctuations
carried respectively by the quantum probability density $f$ and the phase
function $S$. The average quadratic fluctuations $\left\langle \left( \Delta
^{(1)}p_{i}\right) ^{2}\right\rangle $ afford a straightforward
interpretation in terms of classical statistical\ mechanics, by introducing
\ the notions of \emph{\ quantum directional temperature} $T_{QM,i}(t)$ and
\emph{quantum temperature }$T_{QM}(t),$ defined respectively:%
\begin{equation}
mT_{QM,i}(t)\equiv \left\langle \left( \Delta ^{(1)}p_{i}\right)
^{2}\right\rangle ,  \label{Ti}
\end{equation}%
\begin{equation}
T_{QM}(t)=\frac{1}{3}\sum\limits_{i=1,2,3}T_{QM,i}(t),  \label{TQ}
\end{equation}%
which can be viewed as \emph{constitutive equations} for $T_{QM,i}(t)$ and $%
T_{QM}(t).$ We notice that if $\Omega \equiv
\mathbb{R}
^{3N}$ and $\psi (\mathbf{r},t)$ is dynamically consistent \cite%
{Gustafson-Sigal2005}, necessarily it must result $T_{QM}(t)>T_{QM,i}(t)>0$
in $\overline{\Omega }.$ In the remainder we shall assume, however, that it
results $T_{QM}(t)>0$ for all $\in I$ also in the case in which $\Omega $ is
a bounded set. \ As a consequence the following \emph{modified Heisenberg
inequality} holds%
\begin{equation}
\left\langle \left( \Delta r_{i}\right) ^{2}\right\rangle \left[
mT_{QM,i}(t)+\left\langle \left( \Delta ^{(2)}p_{i}\right) ^{2}\right\rangle %
\right] \geq \frac{\hbar ^{2}}{4}.  \label{modified Heisenber inequalities}
\end{equation}%
A particular case is the one in which the \emph{condition of (quantum
temperature) isotropy}
\begin{equation}
T_{QM,i}(t)=T_{QM}(t)  \label{condition of isotropy}
\end{equation}%
holds identically for $i=1,2,3$. In the following sections for greater
generality we shall require, however,
\begin{equation}
T_{QM,i}(t)\neq T_{QM,j}(t),  \label{condition of anisotropy}
\end{equation}%
for $i\neq j$ (with $i,j=1,2,3$). In fact, generally for arbitrary quantum
systems Eq.(\ref{condition of isotropy}) cannot be assumed to hold.

\section{V. Problem: the search of an inverse kinetic theory for NRQM}

\subsection{A. Basic assumptions, the kinetic correspondence principle}

The form of the quantum fluid equations (\ref{Eq.10}) and (\ref{eq.11c})
suggests that they can be obtained as moment equations of a continuous
inverse kinetic theory, analogous to that developed recently for the
incompressible Navier-Stokes equation [see ET].

Let us now pose the problem of searching an inverse kinetic theory for the
Schr\"{o}dinger equation, i.e., a kinetic theory yielding exactly, by means
of suitable moment equations, the quantum hydrodynamic equations. The theory
must hold for arbitrary (and suitably smooth) initial and boundary
conditions both for the wavefunction and the kinetic probability density. In
particular, this involves also the search of a possible underlying classical
dynamical system which determines uniquely the time-evolution of the quantum
system.

In the sequel let us consider, without loss of generality, the case of
one-body quantum systems; the theory here developed is applicable, in fact,
with minor changes also for systems with $N>1$ particles. For definiteness,
let us assume that the quantum fluid fields $\left\{ f(\mathbf{\mathbf{r}}%
,t),\mathbf{V,}T_{QM,i},\text{for }i=1,2,3,\right\} $ are respectively
solutions of the GI-QHE problem [i.e., Eqs. (\ref{Eq.10}), (\ref{eq.11c}), (%
\ref{initial condition for f}), (\ref{initial condition for V-1}), (\ref{BS
f -PRESCRIBED}) - (\ref{BC V - PRESCRIBED -1})] and imposing also the
constitutive equation (\ref{Ti}). To restrict the class of possible kinetic
models, following the approach of ET, let us introduce\ a probability
density $g(\mathbf{x},t),$ with $\mathbf{x=}\left( \mathbf{r,v}\right) ,$
defined in the phase space $\overline{\Gamma }=\overline{\Omega }\times U$
(where $U\equiv
\mathbb{R}
^{3N})$ and assume that it belong to the functional class $\left\{ g(\mathbf{%
x},t)\right\} $ of real functions which satisfy the following properties 1-4
(denoted together as \emph{Assumptions \#1}); \ more precisely, it is
assumed that $g(\mathbf{x},t)$

\begin{enumerate}
\item is non-negative and continuous in $\overline{\Gamma }\times I,$ in
particular, is strictly positive and of class $C^{(k,2)},$ with $k\geq 3,$
in $\Gamma \times I;$

\item $\forall \left( \mathbf{r,}t\right) \in \overline{\Omega }\times I$
admits the velocity moments $M_{X}\left[ g\right] \equiv \int_{U}d\mathbf{vX}%
g,$ with $\mathbf{X}(\mathbf{\mathbf{r,v},}t)=1,\mathbf{v,}u_{i}^{2}$ (for $%
i=1,2,3$)$,\mathbf{uu,u}u^{2},\ln g,$ where $\mathbf{v,}$ $\mathbf{u=v-V}$
are respectively the kinetic and the relative kinetic velocities and $%
u_{i}=v_{i}-V_{i}$ (for $i=1,2,3$) are the orthogonal Cartesian components
of $\mathbf{u}$ defined with respect to an arbitrary inertial reference
frame;

\item admits the velocity moments $\mathbf{X}(\mathbf{\mathbf{r,v},}t)=1,%
\mathbf{v,}u_{i}^{2}$ (for $i=1,2,3$). The latter are prescribed by imposing
a suitable set of constraint equations,to be denoted as \emph{kinetic
correspondence principle,} which relate the quantum fluid fields and the
corresponding kinetic moments.\emph{\ }For this purpose\emph{\ }the
following equations are assumed to hold identically, respectively in $%
\overline{\Omega }\times I,$
\begin{eqnarray}
&&f=M_{1}\left[ g\right] \equiv \int_{U}d\mathbf{v}g(\mathbf{\mathbf{r,v},}%
t),  \label{moment - 1A} \\
&&M_{2}\left[ g\right] \equiv \frac{1}{f(\mathbf{r},t)}\int_{U}d\mathbf{vv}g(%
\mathbf{\mathbf{r,v},}t)=  \label{moment -2A} \\
&=&\mathbf{V}(\mathbf{\mathbf{\mathbf{r}}},t\mathbf{),}  \notag
\end{eqnarray}%
and in $I$ for $i=1,2,3$%
\begin{eqnarray}
&&\left. T_{i}(t)\equiv M_{3i}\left[ g\right] \equiv \right.
\label{moment -3A} \\
&\equiv &\frac{1}{f(\mathbf{r},t)}\int_{U}d\mathbf{v}mu_{i}^{2}g(\mathbf{%
\mathbf{r,v},}t)=T_{QM,i}(t)>0.  \notag
\end{eqnarray}%
Consistently with (\ref{condition of anisotropy}), we shall generally
consider $T_{i}\neq T_{j}$ for $i\neq j$ (with $i,j=1,2,3$). Here the
moments $T_{i}\equiv M_{3i}\left[ g\right] $ (for $i=1,2,3$) and $T=\left(
T_{1}+T_{2}+T_{3}\right) /3$ are denoted respectively the \emph{(quantum)\
kinetic directional temperatures} and the \emph{(quantum) kinetic temperature%
}.

\item Finally, let us impose an appropriate regularity condition for the
fluid fields and the quantum force $\mathbf{F(r,}t\mathbf{)}$. In
particular, besides imposing that the fluid fields $\left\{ f,\mathbf{V,}%
T_{QM,i},\text{for }i=1,2,3,\right\} $ are solutions of the GI-QHE problem
and satisfy the constitutive equations (\ref{Ti}), let us require that they
belong to the functional settings:
\end{enumerate}

\begin{equation}
\left\{
\begin{array}{ll}
f\mathbf{,\mathbf{V,}}T_{i}(\mathbf{\mathbf{r,}}t)\in C^{(k,2)}(\Omega
\times I), &  \\
T_{i}(\mathbf{r,}t\mathbf{),}f(\mathbf{r,}t)>0, &  \\
f\mathbf{,\mathbf{V,}}T_{i}(\mathbf{\mathbf{r,}}t)\in C^{(0)}(\overline{%
\Omega }\times I), &  \\
\mathbf{F(r,}t\mathbf{)}\in C^{(k,2)}(\Omega \times I), &
\end{array}%
\right.  \label{minimal functional setting - 0}
\end{equation}%
\emph{\ }with $k\geq 3.$

The constraint provided by Eq.(\ref{moment -3A}) implies that the kinetic
directional temperatures $T_{i}$\ (for $i=1,2,3$) are assumed
position-independent. This assumption, although consistent with the
definition of the quantum directional temperatures given above [see Eq.(\ref%
{Ti})], may in principle be avoided (see related discussion in Appendix B
and at the end of Sec.6).

Furthermore, let us require that in the open set $\Gamma =\Omega \times U$
the probability density $g(\mathbf{\mathbf{r,v},}t)$ satisfies a Vlasov-type
kinetic equation of the form (\emph{Assumption \#2})
\begin{equation}
\emph{L}g(\mathbf{x,}t)=0  \label{inverse kinetic equation -A}
\end{equation}%
(\emph{inverse kinetic equation}), where \emph{L} is the Vlasov streaming
operator
\begin{equation}
L=\frac{\partial }{\partial t}+\frac{\partial }{\partial \mathbf{x}}\cdot
\left( \mathbf{X}\right) \equiv \frac{d}{dt}+\frac{\partial }{\partial
\mathbf{v}}\cdot \left( \frac{\mathbf{K}}{m}\right)
\label{streaming operator}
\end{equation}%
and$\frac{d}{dt}=\frac{\partial }{\partial t}+\mathbf{v\cdot \nabla }+\frac{%
\mathbf{K}}{m}\cdot \frac{\partial }{\partial \mathbf{v}}$ is the Lagrangian
derivative. Here $\mathbf{x=(r,v)}$ and $\mathbf{X=}\left\{ \mathbf{v,}\frac{%
1}{m}\mathbf{K}\right\} ,$ where $\mathbf{K}(\mathbf{x,}t),$ to be denoted
as \emph{mean field force,} is a suitably smooth real vector field.

Since the correspondence principle defined by Eqs.(\ref{moment - 1A}),(\ref%
{moment -2A}) and (\ref{moment -3A}) must hold identically in the open set $%
\overline{\Omega }\times I$ it follows that the moment equations for $M_{1}%
\left[ g\right] ,M_{2}\left[ g\right] $ and $M_{3i}\left[ g\right] ,$ for $%
i=1,2,3,$ must necessarily coincide identically, respectively, with the
quantum hydrodynamic equations (\ref{Eq.10}) and (\ref{eq.11c}) and the
constitutive equation for the directional temperature (\ref{Ti}). In
addition, it is obvious these moment equations must hold also for arbitrary
quantum fluid fields satisfying assumptions (\ref{minimal functional setting
- 0}). Such implications will be discussed in detail below (see THM. 1 and
2).

\subsection{B. The phase-space Schr\"{o}dinger dynamical system}

Let us remark that the kinetic equation (\ref{inverse kinetic equation -A})
determines uniquely the time evolution of the kinetic distribution function $%
g(\mathbf{x,}t)$ in the whole extended phase-space $\overline{\Gamma }\times
I$ and consequently prescribes uniquely also the quantum fluid fields in the
set $\overline{\Omega }\times I.$ In fact, it can also be cast,
respectively, in the equivalent Lagrangian and integral forms

\begin{equation}
\frac{d}{dt}g(\mathbf{x}(t)\mathbf{,}t)=-g(\mathbf{x}(t)\mathbf{,}t)\frac{%
\partial }{\partial \mathbf{v}(t)}\cdot \frac{\mathbf{K}\left( \mathbf{x}(t)%
\mathbf{,}t\right) }{m},  \label{Lagrangian equation}
\end{equation}%
\begin{equation}
J(\mathbf{x}(t)\mathbf{,}t)g(\mathbf{x}(t)\mathbf{,}t)=g(\mathbf{x}%
_{o},t_{o}),  \label{integral inverse kinetic equation}
\end{equation}%
where $g(\mathbf{x}_{o},t_{o})$ is the initial kinetic distribution function
and the curves $\left\{ \mathbf{x}(t),t\in I\right\} $ define suitable
phase-space Lagrangian trajectories. Moreover, the map
\begin{equation}
\gamma _{\mathbf{(x}_{o},t_{o}\mathbf{)}}:\mathbf{x}_{o}\rightarrow \mathbf{x%
}(t)=\chi (\mathbf{x}_{o},t_{o},t)  \label{flow-A}
\end{equation}%
is the flow generated by the initial-value problem
\begin{eqnarray}
&&\left. \frac{d}{dt}\mathbf{x}\mathbf{=X}(\mathbf{x,}t)\right. ,
\label{equation} \\
&&\mathbf{x}(t_{o})=\mathbf{x}_{o},  \label{initial condition}
\end{eqnarray}%
and
\begin{equation}
J(\mathbf{x}(t)\mathbf{,}t)=\exp \left\{ \int\limits_{t_{o}}^{t}dt^{\prime }%
\frac{\partial }{\partial \mathbf{v}(t^{\prime }\mathbf{)}}\cdot \frac{%
\mathbf{K}\left( \mathbf{x}(t^{\prime })\mathbf{,}t^{\prime }\right) }{m}%
\right\}  \label{Liouville theorem}
\end{equation}%
is its Jacobian

Here we shall prove that, due to continuity of the kinetic distribution
function and of the Schr\"{o}dinger dynamical system, this equation holds
identically in the closure $\overline{\Gamma }\times I,$ except possibly in
the nodes $\mathbf{r}_{n}\in \delta \Omega ,$\emph{\ }where $f(\mathbf{r}%
_{n},t)=0$. However, by construction, the limit
\begin{equation}
\lim_{\mathbf{r}(t)\rightarrow \mathbf{r}_{\delta }}J(\mathbf{x}(t)\mathbf{,}%
t)g(\mathbf{x}(t)\mathbf{,}t)=g(\mathbf{x}_{o},t_{o})
\label{limit of the inverse integral kinetic equation}
\end{equation}%
exists for all $\mathbf{r}_{\delta }\in \delta \Omega ,$ including all
nodes. Instead, one can prove that the limit $\lim_{\mathbf{r}(t)\rightarrow
\mathbf{r}_{\delta }}J(\mathbf{x}(t)\mathbf{,}t)$ does not exist if $\mathbf{%
r}_{\delta }$ is a node. In particular, by suitable definition of the vector
field $\mathbf{K}(\mathbf{x,}t),$ we shall prove that $\chi (\mathbf{x}%
_{o},t_{o},t)$ results suitably regular so that the set of maps (\ref{flow-A}%
) generates a (generally non-conservative) classical dynamical system $%
\left\{ \gamma _{\mathbf{(x}_{o},t_{o}\mathbf{)}}\right\} $, to be denoted
as \emph{phase-space} \emph{Schr\"{o}dinger dynamical system}.

These include in particular - besides Assumptions \#1 and \#2 - the
hypothesis that the kinetic distribution function $g(\mathbf{x}_{o},t_{o})$
(and its initial condition $g(\mathbf{x}_{o},t_{o})$) is suitably smooth in
the whole set $\Gamma \times I$. For example, we let us require that $%
\mathbf{X}(\mathbf{x,}t)\in C^{(k,2)}(\Gamma \times I)$ and $g(\mathbf{x}%
_{o},t_{o})\in $ $C^{(k,2)}(\Gamma \times I),$ with $k\geq 3.$ Then it
follows that $\chi (\mathbf{x}_{o},t_{o},t)$ is a diffeomorphism. of class $%
C^{(k,2,2)}(\Gamma \times I\times I)$ and $g(\mathbf{x}(t)\mathbf{,}t)\in $ $%
C^{(k,2)}(\Gamma \times I),$ with $k\geq 3$ while, so that the moments $%
\left\{ f,\mathbf{V,}T_{i},i=1,2,3\right\} $ are necessarily of class $%
C^{(k,2)}(\Omega \times I)$ (\emph{Assumption \#3}).

Manifestly the Schr\"{o}dinger dynamical system, if it exists, provides a
deterministic description of quantum mechanics since it advances in time
both the kinetic probability density and the quantum fluid fields\emph{\ }$%
\left\{ f(\mathbf{\mathbf{r}},t),\mathbf{V}\right\} $. Thus, a fundamental
issue is the question of the existence of a vector field $\mathbf{K}(\mathbf{%
\mathbf{r,v},}t)$ which satisfies the minimal assumptions indicated above.
To be more specific, however, it is convenient to further specify the
mathematical model imposing an appropriate set of assumptions (denoted
together as \emph{Assumptions \#4}), which include the following ones [ET]:

\begin{enumerate}
\item \emph{the kinetic equation admits local Maxwellian kinetic equilibria
for arbitrary kinetic moments and quantum fluid fields which satisfy the
correspondence principle (\ref{moment - 1A}),(\ref{moment -2A}),(\ref{moment
-3A}) and the regularity requirements (\ref{minimal functional setting - 0});%
}

\item \emph{The vector field }$\mathbf{X(x,}t)$ \emph{is prescribed in such
a way that it depends, besides on }$\mathbf{x},$\emph{\ only on the fluid
fields and suitable differential operators acting on them;}

\item \emph{the kinetic distribution function satisfies appropriate boundary
conditions for the kinetic distribution function; }

\item \emph{the Heisenberg theorem is satisfied identically.}
\end{enumerate}

To complete the specification of the inverse kinetic equation, however, it
must be supplemented with suitable initial and boundary conditions. In
particular since the kinetic correspondence principle must hold both on the
boundary $\delta \Omega $ and at the initial time $t_{o},$ this means that
on the boundary $\delta \Gamma $ (which includes $\delta \Omega $)
bounce-back boundary conditions are imposed on the kinetic distribution
function \cite{Ellero2005}. \

\subsection{C. Bounce-back boundary conditions}

We intend to define boundary conditions for the kinetic distribution
function $g(\mathbf{x,}t)$ which are consistent with the Dirichlet boundary
condition defined on the boundary $\delta \Omega $ for the quantum fluid
fields. Denoting $\mathbf{V}_{w}(\mathbf{r}_{\delta }(t),t)=\frac{d}{dt}%
\mathbf{r}_{\delta }(t)$ the velocity of the point of the boundary
determined by the vector $\mathbf{r}_{\delta }(t)\in \delta \Omega $ and
assuming $\left\vert \mathbf{v-V_{w}}\right\vert \neq 0,$ let us introduce
the unit vector%
\begin{equation}
\mathbf{b}=\sigma \frac{\mathbf{v-V}_{w}}{\left\vert \mathbf{v-V_{w}}%
\right\vert }  \label{unit vector b}
\end{equation}%
and the variable\
\begin{equation}
\xi \equiv \left[ \mathbf{v-V_{w}}\right] \cdot \mathbf{b.}
\label{papameter ki}
\end{equation}%
Here $\sigma =\pm 1$ and its sign is defined so that when $\mathbf{b}$ is a
vector applied at the position $\mathbf{r}_{\delta }$ it is always oriented
inward with respect to the domain $\Omega .$ For sake of definiteness, we
shall assume that $\delta \Omega $ is a piece-wise regular surface and that
the vector $\mathbf{b}$ belongs to the open tangent cone to $\delta \Omega $
in $\mathbf{r}_{n}$ which is oriented inward with respect to $\Omega $.
Then, at an arbitrary position $\mathbf{r}_{\delta }\in \delta \Omega ,$ the
sign of the variable $\xi $ determines \textit{incoming and outgoing
velocity subdomains, }defined, respectively, as the subdomains of velocity
space $U$ for which $\xi <0$ and $\xi >0$. Therefore $\left. g(\mathbf{r}%
_{\delta }\mathbf{,v,}t)\right\vert _{\xi <0}$ and $\left. g(\mathbf{r}%
_{\delta }\mathbf{,v,}t)\right\vert _{\xi >0}$ denote the incoming and
outgoing kinetic distribution functions at position $\mathbf{r}_{\delta }$.
These notations permit us to define properly the boundary conditions for the
kinetic distribution function on $\delta \Omega .$

For any boundary $\delta \Omega ,$position $\mathbf{r}_{n}\in \delta \Omega
_{i}$ and for all nonvanishing vectors $\mathbf{v-V_{w}}$ taking at $\mathbf{%
r}_{\delta }$ the directions specified above, we impose, respectively, for $%
\xi >0$ and $\xi <0$ $\ $the boundary conditions for $g(\mathbf{r,v,}t)$ (%
\emph{Assumption \#5})

\begin{equation}
\left. g(\mathbf{r}_{\delta }\mathbf{,v,}t)\right\vert _{\xi >0}=\left. g(%
\mathbf{r}_{\delta }\mathbf{,2V_{w}-v,}t)\right\vert _{\xi <0},
\label{boundary condifion for f (a)}
\end{equation}%
\begin{equation}
\left. g(\mathbf{r}_{\delta }\mathbf{,v,}t)\right\vert _{\xi <0}=\left. g(%
\mathbf{r}_{\delta }\mathbf{,2V_{w}-v,}t)\right\vert _{\xi >0};
\label{boundary condifion for f (b)}
\end{equation}%
to be denoted as \emph{bounce-back b.c. for }$\emph{g}(\mathbf{x,}t).$ It is
immediate to prove that they are consistent with Dirichlet boundary
conditions defined by Eq.(\ref{BC V - PRESCRIBED -1}). In fact, there
results:
\begin{eqnarray*}
&&\left. f\mathbf{V}(\mathbf{r}_{\delta },t)=\int_{U}d\mathbf{vv}g(\mathbf{r}%
_{\delta }\mathbf{,v,}t)=\right. \\
&&\left. =\int_{U}^{\xi <0}d\mathbf{vv}g(\mathbf{r}_{\delta }\mathbf{,v,}%
t)+\int_{U}^{\xi >0}d\mathbf{vv}g(\mathbf{r}_{\delta }\mathbf{,v,}t)=\right.
\end{eqnarray*}%
\begin{equation}
=\frac{1}{2}\int_{U}^{\xi <0}d\mathbf{vv}\left\{ g(\mathbf{r}_{\delta }%
\mathbf{,v,}t)+g(\mathbf{r}_{\delta }\mathbf{,}2\mathbf{V_{w}-v,}t)\right\} +
\end{equation}%
\begin{equation*}
+\frac{1}{2}\int_{U}^{\xi >0}d\mathbf{vv}\left\{ g(\mathbf{r}_{\delta }%
\mathbf{,v,}t)+g(\mathbf{r}_{\delta }\mathbf{,}2\mathbf{V_{w}-v,}t)\right\} .
\end{equation*}%
Thanks to the identities
\begin{equation}
\frac{1}{2}\int_{U}d\mathbf{vv}g(\mathbf{r}_{\delta }\mathbf{,2V_{w}-v,}t)=%
\left[ \mathbf{V_{w}-}\frac{1}{2}\mathbf{V(r}_{\delta },t)\right] f
\end{equation}%
\begin{equation}
\int_{U}d\mathbf{vv}g(\mathbf{r}_{\delta }\mathbf{,v,}t)=\mathbf{V(r}%
_{\delta },t)f
\end{equation}%
it follows that:
\begin{equation}
\mathbf{V}_{w}(\mathbf{r}_{n},t)\mathbf{=}\frac{1}{f}\int_{U}d\mathbf{vv}g(%
\mathbf{r}_{n}\mathbf{,v,}t)=\mathbf{V}(\mathbf{r}_{n},t).  \label{no-slip}
\end{equation}
We impose, furthermore, the Dirichlet boundary condition (\emph{Assumption
\#6})%
\begin{equation}
\int_{U}d\mathbf{v}g(\mathbf{r}_{\delta }\mathbf{,v,}t)=f_{w}(\mathbf{r}%
_{\delta },t),  \label{boundary condifion for f (c)}
\end{equation}%
where $f_{w}(\mathbf{r}_{\delta },t)$ is the prescribed probability density
on the boundary $\delta \Omega $ defined by (\ref{BS f -PRESCRIBED}).

\section{VI. Construction of the mean-field force $\mathbf{K}$}

\subsection{A. Case of the generalized Maxwellian distribution}

It is well known that the principle of entropy maximization (PEM) \cite{von
Neumann,Jaynes1957,Guiasu1987} can be invoked to determine uniquely, at a
prescribed initial time $t_{o}\in I,$ the initial condition for the kinetic
distribution function $g(\mathbf{x},t).$ This requires that the functional
class $\left\{ g(\mathbf{x},t)\right\} $ must be suitably specified; in
particular PEM can be used to determine the initial condition $g(\mathbf{x}%
,t_{o})$ which corresponds to the requirement that quantum fluid fields ( $%
\left\{ f(\mathbf{\mathbf{r}},t),\mathbf{V,}T_{QM,i},\text{for }%
i=1,2,3,\right\} $ are uniquely prescribed at the initial time $t=t_{o}$. In
such a case the kinetic correspondence principle determines uniquely the
kinetic distribution. This reads necessarily (for $t=t_{o}$)

\begin{equation}
g_{M}(\mathbf{r,v},t)=f(\mathbf{\mathbf{r}},t)\frac{1}{\pi
^{3/2}v_{th1}v_{th2}v_{th3}}\exp \left\{ -x_{i}x_{i}\right\}
\label{Maxwellian}
\end{equation}%
(\emph{generalized Maxwellian distribution}), where in the exponential the
sum is understood on repeated indexes $(i=1,2,3)$. Here the notation is
analogous to ET, thus for $i=1,2,3,$ $v_{thi}=\sqrt{2T_{i}(\mathbf{\mathbf{r}%
},t)/m}$ and $x_{i}=u_{i}/v_{thi}$.

Since the initial time $t_{o}$ is arbitrary, it is natural to assume that $%
g_{M}(\mathbf{r,v},t)$ results identically ($\forall \left( \mathbf{x,}%
t\right) \in \Gamma \times I$) a particular solution of the inverse kinetic
equation (\ref{inverse kinetic equation -A}). In such a case, invoking
Assumptions \#1-\#6 (and in particular the hypothesis that the kinetic
directional temperatures can only depend on time), the mean-field force $%
\mathbf{K}(g_{M})$ results necessarily \emph{(Assumption \#7)}

\begin{equation}
\mathbf{K}(g_{M})=\mathbf{K}_{0}(g_{M})+\mathbf{K}_{1}(g_{M}),
\label{F Maxwellian}
\end{equation}%
with
\begin{eqnarray}
\mathbf{K}_{0}(g_{M}) &=&\mathbf{F}(\mathbf{r,}t)+\frac{1}{f}\frac{m}{2}%
v_{thi}^{2}\widehat{\mathbf{e}}_{i}\widehat{\mathbf{e}}_{i}\cdot \nabla f=
\label{F0 Maxwellian case} \\
&=&\mathbf{F}(\mathbf{r,}t)+\frac{1}{f}\nabla p\mathbf{,}  \notag
\end{eqnarray}%
\begin{equation}
\mathbf{K}_{1}(g_{M})=m\mathbf{u}\cdot \nabla \mathbf{V+}\frac{m}{2}u_{i}%
\widehat{\mathbf{e}}_{i}\frac{\partial }{\partial t}\ln T_{i},
\label{F1 Maxwellian case}
\end{equation}%
where the sum is understood on repeated indexes and $p=fT$ denotes the
kinetic scalar pressure. Here, $\mathbf{K}_{0}(g_{M})$ and $\mathbf{K}%
_{1}(g_{M})$ have been distinguished for being, respectively, constant and
velocity dependent. In particular, $\mathbf{K}_{0}(g_{M})$ contains, besides
the quantum force $\mathbf{F}(\mathbf{r,}t),$ a corrective term ("pressure
term") which depends explicitly only the logarithmic gradient of $f$;
instead $\mathbf{K}_{1}(g_{M})$ contains a "convective term", proportional
to $\nabla \mathbf{V}$ and a contribution proportional to the logarithmic
time derivatives of the directional temperatures. For the sake of reference,
the more general case in which the kinetic directional temperature are taken
as spatially non-uniform is reported in Appendix B.

Let us now examine the main implications which stem, in the particular case $%
g=g_{M}$, from positions (\ref{F Maxwellian})-(\ref{F1 Maxwellian case}) and
assumptions \#1-\#7. \ We first notice that the Schr\"{o}dinger dynamical
system generated by $\mathbf{K}(g_{M}),$ defined as the solution of the
initial value-problem (\ref{equation}),(\ref{initial condition}), exists and
is unique in the whole extended phase-space $\Gamma \times I.$ This property
is manifestly assured by assumption, thanks to to previous definition of $%
\mathbf{K}(g_{M})$ and the regularity properties of the quantum fluid fields
(Assumption \#1). \ In the same set the kinetic distribution function $g_{M}(%
\mathbf{r,v},t)$ is, by construction, a particular solution, i.e., exists,
is unique and results strictly positive. In fact, this property holds if and
only if the quantum fluid fields are solutions of the GI-QHE problem and -
in validity of Eqs.(\ref{F Maxwellian})-(\ref{F1 Maxwellian case}) - if the
kinetic directional temperatures are assumed to be only functions of time,
i.e., $T_{i}(t)$ ($i=1,2,3$). \ In addition, by continuity, the kinetic
distribution function $g_{M}(\mathbf{r,v},t)$ (and its moments) are uniquely
defined also on the boundary set $\delta \Omega $ and hence, in particular,
in the nodes. Finally, it is immediate to prove that Eq.(\ref{inverse
kinetic equation -A}) is also an inverse kinetic equation for the quantum
hydrodynamic equations. Indeed its moment equations, evaluated with respect
to the weight functions $G(\mathbf{x,}t)=1,\mathbf{v,}$ coincide identically
(in $\Omega \times I$) with the Eqs.(\ref{Eq.10}),(\ref{eq.11c}). As a
consequence, the following theorem holds for Maxwellian kinetic
distributions of the type (\ref{Maxwellian}):

\textbf{THEOREM 1 -- Generalized Maxwellian solution of the inverse kinetic
equation}

\emph{Besides the validity of Assumptions \#1-\#7, let us require that:}

\emph{1) the kinetic distribution function fulfills Eq.(\ref{Maxwellian}) at
least at the initial time }$t_{o}\in I;$

\emph{2) the mean-field force }$\mathbf{K}$\emph{\ can depend functionally
but not explicitly on} ${g}_{M}$; \emph{moreover }$K_{0}(g_{M})$\emph{\ and }%
$K_{1}(g_{M})$\emph{\ can only depend on }$\left\{ f(\mathbf{\mathbf{r}},t),%
\mathbf{V,}T_{i},\text{\emph{for} }i=1,2,3,\right\} $.

\emph{Then it follows that:}

\emph{A) }$\forall \left( \mathbf{x,}t\right) \in \overline{\Gamma }\times I$
\emph{the generalized Maxwellian distribution (\ref{Maxwellian}) is } \emph{%
a particular solution of the inverse kinetic equation (\ref{inverse kinetic
equation -A}) if and only if the mean field force }$\mathbf{K}$\emph{\ \ has
the form defined by Eqs.(\ref{F Maxwellian}), (\ref{F0 Maxwellian case}) and
(\ref{F1 Maxwellian case});}

\emph{B) the Schr\"{o}dinger dynamical system exists, is unique and is
continuous in }$\overline{\Gamma }\times I,$\emph{\ except in the nodes,
i.e., when }$\mathbf{r}(t)=\mathbf{r}_{n},$ \emph{with} $f(\mathbf{r}(t)%
\mathbf{,}t)=0.$ \emph{Moreover, it is \ }$C^{(k,2)}(\Gamma \times I)$ \emph{%
with} $k\geq 3.$ \emph{This result holds for arbitrary quantum dynamical
systems and both for isotropic and non-isotropic quantum directional
temperatures;}

\emph{C) } \emph{the Jacobian of the phase-flow} $\mathbf{x}_{o}\rightarrow
\mathbf{x\equiv x}(t)=\chi (\mathbf{x}_{o},t_{o},t),$ \emph{generated by the
initial value problem (\ref{equation}),(\ref{initial condition}), is defined
}$\forall \left( \mathbf{x,}t\right) \in \overline{\Gamma }\times I,$ \emph{%
except in the nodes. There results in such cases}
\begin{equation}
K_{1}(\mathbf{r}_{o}\mathbf{,}t_{o})f(\mathbf{r}(t)\mathbf{,}t)\exp \left\{
-x_{i}(t)x_{i}(t)\right\} \neq 0  \label{non-zero condition}
\end{equation}%
\emph{so that }$J(\mathbf{x}(t),t)$\emph{\ reads :}%
\begin{equation}
J(\mathbf{x}(t),t)=\frac{K_{1}(t)f((\mathbf{r}_{o}\mathbf{,}t_{o}))\exp
\left\{ -x_{io}x_{io}\right\} }{K_{1}(t_{o})f(\mathbf{r}(t)\mathbf{,}t)\exp
\left\{ -x_{i}(t)x_{i}(t)\right\} },  \label{Jacobian}
\end{equation}%
\emph{where} $K_{1}=\left( T_{1}T_{2}T_{3}\right) ^{1/2},$ $%
x_{oi}=u_{oi}/v_{th,i}(t_{o})$, $\mathbf{u}_{o}=\mathbf{v}_{o}\mathbf{-V(r}%
_{n},t_{o}\mathbf{)}$, $x_{i}(t)=u_{i}(t)/v_{th,i}(t_{o})\ $ \emph{and} $%
\mathbf{u}(t)=\mathbf{v(}t)\mathbf{-V(r(}t),t\mathbf{)}$\emph{;}

\emph{D) the limit}
\begin{equation}
\left. \lim_{\mathbf{r}(t)\rightarrow \mathbf{r}_{n}}J(\mathbf{x}(t),t)g_{M}(%
\mathbf{x}(t),t)\right.  \label{limit of J-1}
\end{equation}%
\emph{exists and is unique; }

\emph{E) }$\forall (\mathbf{x},t)\in \Gamma \times I,$ \emph{the
velocity-moment equations of the inverse kinetic equation (\ref{inverse
kinetic equation -A}) evaluated for the weight functions} $G(\mathbf{x,}t)=1,%
\mathbf{v}$ \emph{and for} $g=g_{M}$ \emph{coincide with the fluid equations
Eqs. (\ref{Eq.10}),(\ref{eq.11c});}

\emph{F) }$\forall (\mathbf{x},t)\in \Gamma \times I,$ \emph{the moment
equations for the directional kinetic temperatures are satisfied identically.%
}

\emph{PROOF:}

If the mean-field force $\mathbf{K}$ is assumed of the form (\ref{F
Maxwellian}), (\ref{F0 Maxwellian case}) and (\ref{F1 Maxwellian case}) the
proof of A follows by straightforward algebra from Eq.(\ref{inverse kinetic
equation -A}), or in an equivalent way from the integral equation Eq.(\ref%
{integral inverse kinetic equation}). This furnishes also the proof of Eq.(%
\ref{Jacobian}), which follows invoking\ Liouville theorem (\ref{Liouville
theorem}) (C). Moreover, since all moments $f,\mathbf{V,}$ $T_{i}$ ($i=1,2,3$%
) and the quantum force $\mathbf{F(r,}t\mathbf{)}$ by assumption belong to
the functional setting (\ref{minimal functional setting - 0}), the proof of
B is an immediate consequence of the fundamental (existence and uniqueness)
theorem for ordinary differential equations. In particular, it is obvious
that the solution of the Schr\"{o}dinger dynamical system is not defined in
the nodes since the mean-field force $\mathbf{K}$\emph{\ } is not defined
for $f=0.$ Regarding C, the only possible singular behavior of $J(\mathbf{x}%
(t),t)$ can occur either in the nodes $\mathbf{r}_{n}\in $\ $\delta \Omega ,$%
\ i.e., if at some time $t_{1}\in I$ there results $f(\mathbf{r}%
_{n},t_{1})=0,$ or if least one of the directional temperatures $T_{i}$
vanishes$.$ The second possibility is excluded by assumption. Hence $J(%
\mathbf{x}(t),t)$ is not defined at $(\mathbf{x}(t),t)$ only if $\mathbf{r}%
(t)=\mathbf{r}_{n}$ is a node. Nevertheless, it is obvious that the limit (%
\ref{limit of J-1}) exists (D) and moreover that $g_{M}(\mathbf{x},t)$\emph{%
\ }is by construction non negative and is manifestly defined everywhere in $%
\overline{\Omega }\times I$. Finally the moment equations satisfied by the
inverse kinetic equation for $g=g_{M}(\mathbf{r,v},t)$ are straightforward.
It is immediate to prove that the first two moments $G=1,\mathbf{v}$
coincide respectively with the quantum hydrodynamic equations (\ref{Eq.10}),(%
\ref{eq.11c}),\emph{\ }while $G=u_{i}^{2}$ ($i=1,2,3$) yield moment
equations for the directional temperature which are satisfied identically
(E,F).

\subsection{B. Case of the non-Maxwellian distributions}

As in Ref. ET the inverse kinetic theory can be formulated for
non-Maxwellian kinetic distribution functions too.\ In fact, although PEM
determines uniquely (\ref{Maxwellian}), different initial conditions are
conceivable in which the kinetic distribution function $g(\mathbf{r,v}%
,t_{o})=g_{o}(\mathbf{r,v})$ is an arbitrary function, different from (\ref%
{Maxwellian}), but results otherwise, by assumption, suitably smooth,
strictly positive and summable in $\overline{\Gamma }$. In analogy with ET,
a unique definition of $\mathbf{K}(g)$ which yields the correct fluid
equations and satisfies also the constitutive equations can readily be
obtained also in this case. There results \emph{(Assumption \#7b)}%
\begin{equation}
\mathbf{K}(g)=\mathbf{K}_{0}(g)+\mathbf{K}_{1}(g),  \label{force}
\end{equation}%
\begin{equation}
\mathbf{K}_{0}(g)=\mathbf{F}(\mathbf{r,}t)+\frac{1}{f}\mathbf{\nabla \cdot }%
\underline{\underline{\mathbf{\Pi }}},  \label{force -O}
\end{equation}%
\begin{eqnarray}
&&\left. \mathbf{K}_{1}(g)=m\mathbf{u}\cdot \nabla \mathbf{V+}\right.
\label{force-1} \\
&&+\frac{m}{2}u_{i}\widehat{\mathbf{e}}_{i}\left[ \frac{\partial }{\partial t%
}\ln T_{i}+\frac{3}{fT_{i}}\mathbf{\nabla \cdot Q}_{i}\right] ,  \notag
\end{eqnarray}%
where again the sum is understood on repeated indexes. Equations (\ref{force}%
),(\ref{force-1}) hold if the kinetic directional temperatures are assumed
to be only functions of time [$T_{i}(t),$ for $i=1,2,3$]. In this case the
corrective term in the mean-field force $\mathbf{K}_{0}(g)$ depends on the
tensor pressure $\underline{\underline{\mathbf{\Pi }}},$ instead of the
scalar pressure $p,$ while $\mathbf{K}_{1}(g)$ contains an additional term
depending on the relative heat fluxes $\mathbf{Q}_{i}$ ($i=1,2,3$). The
kinetic moments $\underline{\underline{\mathbf{\Pi }}}$\emph{\ }and $\mathbf{%
Q}_{i}$ (for $i=1,2,3$), assumed to exist, are respectively \emph{\ }%
\begin{eqnarray}
\underline{\underline{\mathbf{\Pi }}} &=&\int d\mathbf{vuu}g, \\
\boldsymbol{Q}_{i} &\boldsymbol{=}&\int d\mathbf{v}\frac{1}{3}\mathbf{u}%
u_{i}^{2}g.
\end{eqnarray}%
The mean-field force defined by Eqs.(\ref{force}),(\ref{force -O}),(\ref%
{force-1}) is manifestly consistent with the previous definition when $%
g=g_{M},$ since in such a case there results $\underline{\underline{\mathbf{%
\Pi }}}=fT\underline{\underline{\mathbf{1}}}\emph{\ =}p\underline{\underline{%
\mathbf{1}}}$\emph{\ }and $\mathbf{Q}_{i}\mathbf{=0}$ ($i=1,2,3$).

It is immediate to prove that also in this case the Schr\"{o}dinger
dynamical system generated by $\mathbf{K}(g)$ exists and is unique in the
whole extended phase-space $\Gamma \times I.$ \ In the same set the kinetic
distribution function $g(\mathbf{r,v},t)$ exists, is unique, results
strictly positive and by continuity is uniquely defined also on the boundary
set $\delta \Omega $. Finally, by construction the moments of the inverse
kinetic equation (\ref{inverse kinetic equation -A}), evaluated with respect
to the weight functions $G(\mathbf{x,}t)=1,\mathbf{v,}$ coincide identically
(in $\Omega \times I$) with the quantum hydrodynamic equation Eqs.(\ref%
{Eq.10}),(\ref{eq.11c}). As a consequence, in the case of non-Maxwellian
kinetic distributions the following theorem holds:

\textbf{THEOREM 2 -- Non-Maxwellian solutions of the inverse kinetic equation%
}

\emph{Besides the validity of Assumptions \#1-\#7, with \#7b replacing \#7,
let us require that:}

\emph{1) the kinetic distribution function is a smooth function of class }$%
C^{(k,2)}(\Gamma \times I),$ \emph{with} $k\geq 2,$ \emph{which is suitably
summable in }$\Gamma ;$

\emph{2) the mean-field force }$\mathbf{K}$\emph{\ can depend functionally
but not explicitly on} $g;$ \emph{moreover }$K_{0}(g)$\emph{\ and }$K_{1}(g)$%
\emph{\ depend, besides $\left\{ f(\mathbf{\mathbf{r}},t),\mathbf{V,}T_{i},%
\text{for }i=1,2,3,\right\}$, at most on the higher-order moments } $%
\underline{\underline{\mathbf{\Pi }}}$\emph{\ and }$\mathbf{Q}_{i}$\emph{\ (}%
$i=1,2,3$\emph{). }

\emph{Then it follows that:}

\emph{A) if at a prescribed time }$t\in I$ \emph{and }$\forall \mathbf{x}\in
\overline{\Gamma }$ \emph{the kinetic distribution function }$g(\mathbf{x}%
,t) $\emph{\ is of the form (\ref{Maxwellian}) then it results }$g(\mathbf{x}%
,t)=g_{M}(\mathbf{x},t),$\emph{\ }$\forall \left( \mathbf{x,}t\right) \in
\overline{\Gamma }\times I;$

\emph{B) }$\forall (\mathbf{x},t)\in \Gamma \times I$ \emph{and for
arbitrary smooth }$g(\mathbf{x,}t)$\emph{\ the velocity-moment equations of
the inverse kinetic equation (\ref{inverse kinetic equation -A}) evaluated
for the weight functions} $G(\mathbf{x,}t)=1,\mathbf{v}$ \emph{coincide with
the fluid equations Eqs. (\ref{Eq.10}),(\ref{eq.11c});}

\emph{C) }$\forall (\mathbf{x},t)\in \Gamma \times I,$ \emph{the moment
equations for the directional kinetic temperatures are satisfied identically;%
}

\emph{D) } \emph{the Jacobian of the phase-flow} $\mathbf{x}_{o}\rightarrow
\mathbf{x\equiv x}(t)=\chi (\mathbf{x}_{o},t_{o},t),$ \emph{generated by the
initial value problem (\ref{equation}),(\ref{initial condition}), is defined
\ }$\forall \left( \mathbf{x,}t\right) \in \overline{\Gamma }\times I$ \emph{%
reads in this case:}%
\begin{eqnarray}
&&\left. J(\mathbf{x}(t)\mathbf{,}t)=\frac{K_{1}(\mathbf{r}(t)\mathbf{,}t)f(%
\mathbf{r}_{o}\mathbf{,}t_{o})}{K_{1}(\mathbf{r}_{o}\mathbf{,}t_{o})f(%
\mathbf{r}(t)\mathbf{,}t)}\right.  \label{Jacobian-2} \\
&&\exp \left\{ \int\limits_{t_{o}}^{t}dt^{\prime }G(\mathbf{x}(t^{\prime
}),t^{\prime })\right\}  \notag
\end{eqnarray}%
\emph{where }$K_{1}=\left( T_{1}T_{2}T_{3}\right) ^{1/2}$\emph{\ and}
\begin{equation}
\left. G(\mathbf{x}(t),t)=\frac{1}{f}\mathbf{u\cdot }\nabla f\mathbf{+}\frac{%
1}{2}\frac{1}{fT_{i}}\mathbf{\nabla \cdot Q}_{i}\right. ;
\end{equation}

\emph{E) if }$g(\mathbf{x},t_{o})$\emph{\ is strictly positive }$\forall (%
\mathbf{x})\in \overline{\Gamma },$ \emph{then for all }$\forall (\mathbf{x}%
,t)\in \overline{\Gamma }\times I$,\ $g(\mathbf{x},t)$ \emph{is also
strictly positive};

\emph{F) the Schr\"{o}dinger dynamical system exists and is unique and is
continuous in }$\overline{\Gamma }\times I,$\emph{\ except in the nodes,
i.e., when }$\mathbf{r}(t)=\mathbf{r}_{n},$ \emph{with} $f(\mathbf{r}(t)%
\mathbf{,}t)=0.$ \emph{Moreover it is \ }$C^{(k,2)}(\Gamma \times I)$ \emph{%
with} $k\geq 3.$ \emph{This result holds for arbitrary quantum dynamical
systems and both for isotropic and non-isotropic quantum directional
temperatures.}

\emph{PROOF:}

The proof of A is immediate thanks to THM.1, since by construction $\mathbf{K%
}(g)=\mathbf{K}(g_{M})$. The proof of B and C follows by direct inspection.
In particular, again \ the directional temperatures $T_{i}$ ($i=1,2,3$) are
completely arbitrary, but suitably smooth, time-dependent functions; hence
they can be uniquely determined according to Eq.(\ref{moment -3A}). Eq.(\ref%
{Jacobian-2}) in D follows after straightforward algebra invoking Liouville
theorem (\ref{Liouville theorem}). In particular, once again it follows that
$J(\mathbf{x}(t)\mathbf{,}t)$ is not defined if $\mathbf{r}(t)$ [or $\mathbf{%
r}_{n}$] coincide with a\emph{\ }node $r_{w}\in $\ $\delta \Omega $\ where%
\emph{\ }$f(\mathbf{r}_{n}\mathbf{,}t)=0.$ As a consequence $\forall (%
\mathbf{x},t)\in \Gamma \times I,$ $J(\mathbf{x}(t)\mathbf{,}t)>0.$ This
proves also propositions E and F. Finally, to reach the proof of uniqueness
of the mean-field force $\mathbf{K}(g)$ it is sufficient to prescribe,
besides the assumption that its does not depend explicitly on $g,$ also an
appropriate dependence in terms of higher-order kinetic moments.

An important consequence of the previous theorems is the uniqueness of the
mean-field force $\mathbf{K}(g)$ and hence of the inverse kinetic equation.

\textbf{THEOREM 3 -- Uniqueness theorem for }$\mathbf{K}(g)$\emph{\ }

\emph{In validity of THM.1 and 2 the mean field force, for which A-E hold, }$%
\mathbf{K}(g)$\emph{\ \ is unique. For }$g=g_{M}$\emph{\ it has the form
defined by Eqs.(\ref{F Maxwellian}), (\ref{F0 Maxwellian case}) and (\ref{F1
Maxwellian case}); \ for }$g$ $\neq g_{M}$ \emph{the mean field-force has
the form defined by Eqs. (\ref{force}),(\ref{force -O}) and (\ref{force-1}).
}

\emph{PROOF}

For example, let us prove that in validity of assumption 3) of THM.1 the
mean-field force $\mathbf{K}(g_{M})$ is unique\emph{. }Let us assume that
there is a nonvanishing vector field $\Delta \mathbf{K}$ such that $\mathbf{K%
}^{\prime }=\mathbf{K+}\Delta \mathbf{K}$ is also an admissible mean-field
force. Hence it must be
\begin{equation}
\frac{\partial }{\partial \mathbf{v}}\cdot \left( \Delta \mathbf{K}{g}%
_{M}\right) =0.
\end{equation}%
This means that $\Delta \mathbf{K}{g}_{M}$ has necessarily the form%
\begin{equation}
\Delta \mathbf{K}{g}_{M}\mathbf{=}\frac{\partial A}{\partial \mathbf{v}}%
\times \frac{\partial B}{\partial \mathbf{v}},
\end{equation}%
where $A$ and $B$ are suitably smooth real scalar fields. If both $A$\textbf{%
\ }and $B$ are independent of ${g}_{M}$ then it must be $\Delta \mathbf{%
K\equiv 0.}$ Instead, for example, let us assume that only $A$ depends on ${g%
}_{M}.$ Letting
\begin{equation}
\frac{1}{{g}_{M}}\frac{\partial A({g}_{M},\mathbf{r,v},t)}{\partial \mathbf{v%
}}=\widehat{\mathbf{A}},
\end{equation}%
if follows that $\widehat{\mathbf{A}}$ must depend on $\{g_{M}$ too. Hence,
in order that $\Delta \mathbf{K}$ results independent of ${g}_{M}$ it must
vanish identically. This proves that it must be $\Delta \mathbf{K\equiv 0}$
and hence $\mathbf{K}$ is unique.

Let us briefly comment on these results.

First, we stress that Schr\"{o}dinger dynamical system [defined by the
initial-value problem (\ref{equation}),(\ref{initial condition})] uniquely
generates the time evolution of the kinetic distribution function $g$ and
hence of its moments. The result \ holds not only for generalized Maxwellian
distributions (\ref{Maxwellian}) but also for arbitrary, but otherwise
suitably smooth, kinetic probability densities (THM.2). An interesting
aspect concerns, in particular, the behavior of the kinetic distribution
function and of the Schr\"{o}dinger dynamical system in the nodes, i.e., the
points of the boundary $\delta \Omega $ in which the probability density $f(%
\mathbf{r},t)$\ vanishes. Their occurrence is usually associated to the
possible singular behavior of the wavefunction, i.e., its non-uniqueness
(for a review see \cite{Deotto}). Therefore, the question arises whether the
existence of these roots is reflected in any way in the inverse kinetic
approach (and therefore on the quantum system). It is manifest that in these
points the dynamical system is not defined and hence the Jacobian of the Schr%
\"{o}dinger dynamical system [given by (\ref{Jacobian}) or more generally by
Eq.(\ref{Jacobian-2})] vanishes or is not defined. \ However, in these
points the kinetic probability density can still be uniquely defined, thanks
to continuity. Consequently, the kinetic distribution function $g$ and its
moments, and in particular those corresponding to the relevant quantum fluid
fields, are defined without exceptions in the whole domain $\overline{\Omega
}\times I.$ As a consequence, it follows that \emph{no singularity appears,
i.e., the quantum fluid fields (and } \emph{hence} \emph{the quantum
wavefunction) are unique in} $\overline{\Omega }\times I$.

\subsection{C. The kinetic representation of Heisenberg inequalities}

It is interesting to examine the role of the kinetic directional
temperatures $T_{i}$ ($i=1,2,3$) in the framework of the inverse kinetic
approach and the consequent interpretation of the Heisenberg theorem. For
this purpose, let us analyze the implications due to the kinetic
correspondence principle. It is immediate to prove that the Heisenberg
inequalities (\ref{Heisenber inequalities-2}) can be represented, in terms
of the \emph{kinetic standard deviations} for\emph{\ }position and linear
momentum $\overline{\Delta }r_{i}^{kin},\overline{\Delta }p_{i}^{kin},$ in
the form:%
\begin{equation}
\overline{\Delta }r_{i}^{kin}\overline{\Delta }p_{i}^{kin}\geq \frac{\hbar }{%
2},  \label{Hesenber-inequalities-3}
\end{equation}%
where $\overline{\Delta }r_{i}^{kin}$ and $\overline{\Delta }p_{i}^{kin}$
are defined in terms of appropriate phase-space averages. For this purpose,
introducing the phase-space average $\left\langle \left\langle
A\right\rangle \right\rangle =\int\limits_{\Gamma }d\mathbf{x}g(\mathbf{r,v}%
,t)A(\mathbf{r,v},t),$where $A(\mathbf{r,v},t)$ is an arbitrary summable
phase-space function, let us pose respectively $\overline{\Delta }%
r_{i}^{kin}=\left\langle \left\langle \left( \Delta r_{i}\right)
^{2}\right\rangle \right\rangle ^{1/2},$ $\overline{\Delta }%
p_{i}^{kin}=\left\langle \left\langle \left( \Delta \overline{\overline{p}}%
_{i}^{kin}\right) ^{2}\right\rangle \right\rangle ^{1/2}$ \ (for $i=1,2,3$),
where $\left\langle \left\langle \left( \Delta r_{i}\right)
^{2}\right\rangle \right\rangle $ and $\overline{\Delta }p_{i}^{kin}=\left%
\langle \left\langle \left( \Delta \overline{\overline{p}}_{i}^{kin}\right)
^{2}\right\rangle \right\rangle $ are denoted as \emph{average quadratic
kinetic fluctuations}. \emph{\ }Here $\Delta \mathbf{r}=\mathbf{r-}%
\left\langle \mathbf{r}\right\rangle $ is the position fluctuation, while $%
\Delta \overline{\overline{\mathbf{p}}}^{kin}=\mathbf{p}^{kin}\mathbf{-}%
m\left\langle \mathbf{V}\right\rangle $ and $\mathbf{p}^{kin}=m\mathbf{v}$
are respectively the \emph{kinetic momentum fluctuation,} and the \emph{%
kinetic momentum.} It is immediate to prove that (\ref%
{Hesenber-inequalities-3}) are equivalent to (\ref{Heisenber inequalities-2}%
). In fact, \ in analogy with Eq.(\ref{momentum fluctuations-2}), one finds
that the average quadratic momentum fluctuation $\left\langle \left\langle
\left( \Delta \overline{\overline{p}}_{i}^{kin}\right) ^{2}\right\rangle
\right\rangle $ can be written in the form%
\begin{equation}
\left\langle \left\langle \left( \Delta \overline{\overline{p}}%
_{i}^{kin}\right) ^{2}\right\rangle \right\rangle =\left\langle \left\langle
\left( \Delta p_{i}^{kin}\right) ^{2}\right\rangle \right\rangle
+\left\langle \left( \Delta ^{(2)}p_{i}\right) ^{2}\right\rangle
\label{implication-00}
\end{equation}%
($i=1,2,3$), where $\Delta \mathbf{p}^{kin}$ denotes $\Delta \mathbf{p}%
^{kin}=\mathbf{p}^{kin}\mathbf{-}m\mathbf{V}(\mathbf{r},t)$ and$\
\left\langle \left( \Delta ^{(2)}p_{i}\right) ^{2}\right\rangle $ is given
by Eq.(\ref{part B}). Moreover, by definition, it follows that the momentum
fluctuation $\left\langle \left\langle \left( \Delta p_{i}^{kin}\right)
^{2}\right\rangle \right\rangle $\emph{\ }reads\emph{\ }
\begin{equation}
\left\langle \left\langle \left( \Delta p_{i}^{kin}\right) ^{2}\right\rangle
\right\rangle =mT_{i}(t),  \label{imlication-0}
\end{equation}%
where and $T_{i}(t)$ (for $i=1,2,3$) are the directional kinetic
temperatures. As a consequence, the constraint $T_{i}(t)=T_{QM,i}(t),$\ set
(for $i=1,2,3$) by Eq.(\ref{moment -3A}) of the correspondence principle,
implies
\begin{equation}
\left\langle \left( \Delta ^{(1)}p_{i}\right) ^{2}\right\rangle
=\left\langle \left\langle \left( \Delta p_{i}^{kin}\right)
^{2}\right\rangle \right\rangle .  \label{implication-1}
\end{equation}%
It follows that the Heisenberg inequalities can be interpreted in terms of
kinetic fluctuations, i.e., as constraints between $\left\langle
\left\langle \left( \Delta r_{i}\right) ^{2}\right\rangle \right\rangle $
and $\overline{\Delta }p_{i}^{kin}=\left\langle \left\langle \left( \Delta
\overline{\overline{p}}_{i}^{kin}\right) ^{2}\right\rangle \right\rangle ,$
whereby the quantum observable\emph{\ }$\mathbf{p}=-i\hbar \nabla $ and its
average quadratic quantum fluctuation are replaced \ by by the kinetic
momentum $\mathbf{p}^{kin}=m\mathbf{v}$ and the corresponding average
quadratic kinetic fluctuation. This result is a direct consequence of the
assumption set by Eq.(\ref{moment -3A}) for the directional kinetic
temperatures $T_{i}$ ($i=1,2,3$).

\subsection{D. Generalizations: non-uniqueness}

It is obvious that the present results can be generalized in several ways.
In particular, the definition of the kinetic directional temperatures
remains in principle arbitrary since they do not enter explicitly the
quantum hydrodynamic equations. Thus, for example, it is possible to require
that the functions $T_{i}$ ($i=1,2,3$) are also position-dependent (see
Appendix B).

Due to the arbitrariness of the kinetic temperatures, it follows that there
exist infinite equivalent realizations of the Schr\"{o}dinger dynamical
system and of the associated Lagrangian trajectories $\left\{ \mathbf{x}%
(t),t\in I\right\} $ which yield the same quantum hydrodynamic equations.
Therefore, the unique inverse kinetic theory here presented, which
corresponds to a well-defined set of prescriptions and in particular the
assumption of spatially-constant directional temperatures, is simply one of
the infinite possible mathematical realizations.

A side aspect concerns the so-called uniqueness problem of the deterministic
viewpoint of SQM \cite{Bohm1952a,Holland,Deotto}, i.e., \ the Bohmian
program, of reproducing the predictions of SQM within the framework of
suitable deterministic Lagrangian trajectories. In fact, the Schr\"{o}dinger
dynamical system determined by the initial-value problem (\ref{equation}),(%
\ref{initial condition}) yields in terms of the associated Lagrangian
trajectories $\left\{ \mathbf{x}(t),t\in I\right\} $ a deterministic
description of SQM. Hence, it can also be viewed a phase-space
generalization of Bohmian mechanics \cite{Bohm1952a,Bohm1952b,Bohm1952c}. An
implication of the present theory is that such a program has by no means a
unique solution. In fact, there are infinite equivalent Lagrangian
trajectories determined as solutions of the initial-value problem (\ref%
{equation}),(\ref{initial condition}), which differ only by the choice of
the spatial-dependency assumed for the directional temperatures (see
Appendix B).

\section{VII. Conclusions}

Motivated by the analogy between hydrodynamic description of SQM and
classical fluid dynamics an inverse kinetic theory has been developed for
the quantum hydrodynamic equations. We have shown that, although in
principle infinite solutions to this problem exist (in particular due to the
indeterminacy in the kinetic directional temperatures), the inverse kinetic
theory can be uniquely determined, provided appropriate hypotheses are
introduced. The results presented are relevant for the fluid description of
quantum mechanics and a deeper understanding of the underlying statistical
(in particular, kinetic) descriptions.

For this purpose the full set of gauge-invariant quantum hydrodynamic
equations, including the Heisenberg inequalities, have been related to the
appropriate quantum fluid fields. As a result the notions of quantum
temperature and quantum directional temperatures have been introduced.\ The
present approach has the following main features:

\begin{enumerate}
\item the inverse kinetic equation (\ref{inverse kinetic equation -A}) has
been assumed to be a Vlasov-type kinetic equation;

\item its solution, i.e., the kinetic distribution function has been
required, in particular, to admit kinetic directional temperatures $T_{i}$ ($%
i=1,2,3$) which, consistent with the correspondence principle [defined by
Eqs.(\ref{moment - 1A}),(\ref{moment -2A}) and (\ref{moment -3A})], depend
only on time;

\item the inverse kinetic theory holds for suitably smooth, but otherwise
arbitrary quantum fluid fields, which satisfy the quantum hydrodynamics
equations with appropriate initial-boundary value conditions;

\item by imposing a kinetic correspondence principle, i.e., by identifying
the relevant quantum fluid fields with appropriate kinetic moments, the
quantum hydrodynamic equations are satisfied identically, when are expressed
in terms of the relevant moment equations.

\item theory is non-asymptotic, i.e., the quantum hydrodynamic equations are
satisfied exactly;

\item the moment equations form a complete system of equations (closure
condition).

\item under suitable assumption, the inverse kinetic theory and the
mean-field force which defines the streaming operator are unique.
\end{enumerate}

An interesting result of the theory, relevant for the mathematical
investigation of the Schr\"{o}dinger equation, concerns the discovery of the
underlying dynamical system, i.e., the phase-space Schr\"{o}dinger dynamical
system. We have found that this can be identified with the non-conservative
dynamical system advancing in time the kinetic distribution function and
generated by the kinetic equation itself. The evolution of the fluid fields
is proven to be determined uniquely by this dynamical system. Formally the
Schr\"{o}dinger dynamical system can be interpreted as describing the
dynamics of system of classical "virtual" subquantum particles which
interact with each other only by means the mean-field force $\mathbf{K}$ and
are characterized by a dynamics which fulfills a suitable set of regularity
assumptions.

To conclude, a further interesting feature of the present treatment is its
adoption of the mean field force kinetic model. This permits, in principle,
numerical implementations by means of appropriate algorithms based on the
new inverse kinetic theory [see related discussion in ET].

A side aspect concerns also the uniqueness and regularity of the quantum
fluid fields and of the related quantum wavefunction. We have shown, in
fact, that the kinetic distribution function is smooth in the whole extended
phase space $\Gamma \times I$, while is uniquely defined also in the nodes,
i.e., the "singular" points of configuration space $\overline{\Omega }$
where the quantum probability density $f(\mathbf{r},t)$ vanishes. As a
consequence, the quantum fluid fields are necessarily unique in $\overline{%
\Omega }\times I$ and suitably smooth in $\Omega \times I$.

Finally, have pointed out that the Heisenberg inequalities afford a simple
statistical interpretation, which permits the representation of the quantum
statistical fluctuations of the components of the linear momentum, and
corresponding quantum directional temperatures in terms of statistical
fluctuations of the kinetic momentum and of the kinetic directional
temperatures. \

\section*{Acknowledgments}

Research developed in the framework of MIUR PRIN project \textquotedblleft
Fundamentals of kinetic theory and applications to fluid dynamics,
magnetofluiddynamics and quantum mechanics \textquotedblright \thinspace\
partially supported by GNFM (Gruppo Nazionale di Fisica Matematica) and CMFD
Consortium (Consorzio di Magnetofluidodinamica), Trieste (Italy).

\section{APPENDIX A: Reduced one-particle description of quantum systems}

In SQM the state of a system of $N$ interacting particles is, by assumption,
represented by its $N-$body wavefunction $\psi (\mathbf{r},t),$ with \textit{%
\ }$\mathbf{r}=\left( \mathbf{r}_{1},...,\mathbf{r}_{N}\right) \in \overline{%
\Omega },$ $\mathbf{r}_{j}\in \overline{\Omega }_{j}$\textit{\ }(for $j=1,N$%
) and $t\in I.$ This is defined in the set $\overline{\Omega }\times I,$
where $\Omega $ is the configuration space $\Omega \equiv
\prod\limits_{j=1,N}\Omega _{j},$ $\Omega _{i}\subseteq
\mathbb{R}
^{3}$ (for $j=1,N$) and $I$\ is an open subset of $%
\mathbb{R}
.$ However, since the number of particles forming a quantum dynamical system
is "a priori" arbitrary, also "reduced" quantum descriptions of an $N-$body
system are permitted. These descriptions, however, are not equivalent to the
full $N-$body description based on the $N-$body wavefunction. Thus, it is in
principle possible to obtain a \emph{reduced description} based, for
example, on one-particle wavefunctions, whereby the $N-$body system is
represented by the \emph{reduced vector state}
\begin{equation}
\psi _{R}(\mathbf{r},t)\equiv (\psi _{1}(\mathbf{r}_{1},t),....\psi _{N}(%
\mathbf{r}_{N},t)),  \label{one-particle reduce representation}
\end{equation}%
instead of the single scalar $N-$particle wavefunction $\psi (\mathbf{r},t)$%
. The one-particle wavefunction $\psi _{j}(\mathbf{r}_{j},t)$ - which
prescribes the state of the $j$-th one-particle subsystem - is defined by
means of the integral%
\begin{equation}
c_{(j)}(t)\psi _{j}(\mathbf{r}_{j},t)=L_{j}\psi ^{(N)}(\mathbf{r},t)\equiv
\widehat{\psi }_{j}(\mathbf{r}_{j},t),
\end{equation}%
where $L_{j}$ is the integral operator
\begin{equation}
L_{j}=\int\limits_{\prod\limits_{k=1,N;k\neq j}\Omega _{k}}\frac{d\mathbf{r}%
_{1}..d\mathbf{r}_{N}}{d\mathbf{r}_{j}}.
\end{equation}%
Here $c_{j}(t)$ (for $j=1,N$) are a real functions defined so that there
results identically
\begin{equation}
\int\limits_{\Omega _{j}}d\mathbf{r}_{j}\left\vert \psi _{j}(\mathbf{r}%
_{j},t)\right\vert ^{2}=\frac{1}{c_{j}^{2}(t)}\int\limits_{\Omega _{j}}d%
\mathbf{r}_{j}\left\vert \widehat{\psi }_{j}(\mathbf{r}_{j},t)\right\vert
^{2}=1
\end{equation}%
and $f_{j}=\left\vert \psi _{j}(\mathbf{r}_{j},t)\right\vert ^{2}$ are the
associated one-particle probability densities. The Schr\"{o}dinger equation
for $\psi _{j}(\mathbf{r}_{j},t)$ follows immediately from the $N-$body Schr%
\"{o}dinger equation (\ref{Eq.3b}). Let us assume for definiteness that the $%
N-$body Hamiltonian takes the form%
\begin{eqnarray}
H &=&\sum\limits_{k=1,N}H_{ok}+\sum\limits_{k,m=1,N;k<m}U_{km}(\mathbf{r}%
_{k},\mathbf{r}_{m},t)+ \\
&&+\sum\limits_{k=1,N}U_{0k}(\mathbf{r}_{k},t),
\end{eqnarray}%
where $U_{km}(\mathbf{r}_{k},\mathbf{r}_{m},t)$ and $U_{0k}(\mathbf{r}%
_{k},t) $ are respectively binary and unary interaction potentials.
Introducing the position%
\begin{eqnarray}
&&\left. L_{j}\sum\limits_{k,m=1,N;k<m}U_{km}(\mathbf{r}_{k},\mathbf{r}%
_{m},t)\psi ^{(N)}(\mathbf{r},t)=\right. \\
&=&c_{(j)}(t)U_{1}(\mathbf{r}_{j},t)\psi _{j}(\mathbf{r}_{j},t),
\end{eqnarray}%
it follows
\begin{equation}
L_{j}H\psi =c_{(j)}(t)\left\{ H_{o\left( j\right) }+U_{1}(\mathbf{r}%
_{j},t)+U_{0j}\right\} \psi _{j}(\mathbf{r}_{j},t).
\end{equation}%
Hence, $\psi _{j}(\mathbf{r}_{j},t)$ obeys necessarily the one-particle Schr%
\"{o}dinger equation%
\begin{equation}
i\hbar \frac{\partial }{\partial t}\psi _{j}=H_{\left( j\right) }\psi _{j},
\label{Eq.3BB}
\end{equation}%
where the index $j$ (for $j=1,N$) identifies the particle subsystem (or
\emph{species }index) and%
\begin{equation}
H_{j}=H_{oj}+U_{0j}-i\hbar \frac{\partial }{\partial t}\ln c_{(j)}
\label{one-particle Hamiltonian}
\end{equation}%
is the $j$-th particle Hamiltonian. The reduced one-particle description of
a $N-$body quantum system is, therefore, obtained by means of the vector
wavefunction $\psi (\mathbf{r},t)\equiv \left( \psi _{1},....\psi
_{N}\right) ,$ $\psi _{j}(\mathbf{r}_{j},t),$ for $j=1,N,$ being the
one-particle wavefunctions which obey Eq.(\ref{Eq.3BB})$.$

\section{APPENDIX B: Case of position-dependent directional temperatures}

We notice that the correspondence principle (\ref{moment -3A}) can be
modified by assuming instead $T_{i}=T_{i}(\mathbf{r,}t)$ ($i=1,2,3$) and
imposing, in place of Eq.(\ref{moment -3A}), the constraint equation%
\begin{equation}
\left\langle T_{i}(\mathbf{r,}t)\right\rangle =T_{QM,i}(t),
\label{moment-3AA}
\end{equation}%
with general solution of the form%
\begin{eqnarray}
&&\left. T_{i}(\mathbf{r,}t)=k_{(i)}(\mathbf{r,}t)\left\langle T_{i}(\mathbf{%
r,}t)\right\rangle ,\right.  \label{general solution} \\
&&\left. \left\langle k_{(i)}(\mathbf{r,}t)\right\rangle =1.\right.
\label{constraint -2}
\end{eqnarray}%
The functions $k_{(i)}(\mathbf{r,}t)$ ($i=1,2,3$) which satisfy Eq.(\ref%
{constraint -2}) are manifestly non-unique. In this case it is immediate to
prove that for the generalized Maxwellian solution (\ref{Maxwellian}) the
mean-field $\mathbf{K}(g_{M})$ is obtained by imposing (\ref{F Maxwellian}),(%
\ref{F0 Maxwellian case}) with

\begin{eqnarray}
&&\left. \mathbf{K}_{1}(g_{M})=m\mathbf{u}\cdot \nabla \mathbf{V+}\frac{m}{2}%
u_{i}\widehat{\mathbf{e}}_{i}\frac{D}{Dt}\ln T_{i}-\right.  \notag \\
&&-\frac{m}{2}\widehat{\mathbf{e}}_{i}\widehat{\mathbf{e}}%
_{i}v_{th,i}^{2}\cdot \sum\limits_{j=1,2,3}\nabla \ln T_{j}(x_{j}^{2}-\frac{1%
}{2})-  \label{F1 Maxwellian case BB-1} \\
&&-\frac{m}{2}\widehat{\mathbf{e}}_{i}\widehat{\mathbf{e}}%
_{i}v_{th,i}^{2}\cdot \nabla \ln T_{i}  \notag
\end{eqnarray}%
replacing Eq. (\ref{F1 Maxwellian case}). Here, again and the sum is
understood on repeated indexes. \ The general case in which $g\neq g_{M}$
can be obtained immediately from Eqs.(\ref{F1 Maxwellian case BB-1}) and (%
\ref{force -O}),(\ref{force-1}).

\end{document}